\documentclass[a4paper,fleqn,usenatbib]{mnras}

\usepackage{newtxtext,newtxmath}

\usepackage[T1]{fontenc}
\usepackage{ae,aecompl}

\usepackage{graphicx}	
\usepackage{amsmath}	
\usepackage{amssymb}	

\usepackage{multirow}   
\usepackage{xcolor}     
\usepackage{pdflscape}  

\newcommand{\Teff}{$T_{\text{eff}}$}
\newcommand{\logg}{$\log(g)$}
\newcommand{\loggunits}{$\log(g/\text{cm s$^{-2}$})$}
\newcommand{\drvm}{$\Delta\text{RV}_{\text{max}}$}
\newcommand{\kms}{km s$^{-1}$}
\newcommand{\feh}{[Fe/H]}
\newcommand{\alh}{[$\alpha$/H]}
\newcommand{\alfe}{[$\alpha$/Fe]}
\newcommand{\oh}{[O/H]}
\newcommand{\mgh}{[Mg/H]}
\newcommand{\sih}{[Si/H]}
\newcommand{\msun}{M$_{\odot}$}

\title[Stellar multiplicity in APOGEE]{The Close Binary Fraction as a Function of Stellar Parameters in APOGEE: A Strong Anti-Correlation With $\alpha$ Abundances}

\author[C. Mazzola et al.]{Christine N.\ Mazzola,$^{1}$\thanks{E-mail: cnm37@pitt.edu}
Carles Badenes,$^{1}$
Maxwell Moe,$^{2}$
Sergey E.\ Koposov,$^{3,4}$
\newauthor
Marina Kounkel,$^{5}$
Kaitlin Kratter,$^{2}$
Kevin Covey,$^{5}$
Matthew G.\ Walker,$^{6}$
\newauthor
Todd A.\ Thompson,$^{7}$
Brett Andrews,$^{1}$
Peter E.\ Freeman,$^{8}$
Borja Anguiano,$^{9}$
\newauthor
Joleen K.\ Carlberg,$^{10}$
Nathan M.\ De Lee,$^{11,12}$
Peter M.\ Frinchaboy,$^{13}$
Hannah M.\ Lewis,$^{9}$
\newauthor
Steven Majewski,$^{9}$
David Nidever,$^{14,15}$
Christian Nitschelm,$^{16}$ 
Adrian M.\ Price-Whelan,$^{17}$
\newauthor
Alexandre Roman-Lopes,$^{18}$
Keivan G.\ Stassun,$^{12}$
Nicholas W.\ Troup$^{19}$
\\
$^{1}$Department of Physics and Astronomy and Pittsburgh Particle Physics, Astrophysics and Cosmology Center (PITT PACC),\\ University of Pittsburgh, 3941 O`Hara Street, Pittsburgh, PA 15260, USA\\
$^{2}$Steward Observatory, University of Arizona, 933 North Cherry Avenue, Tucson, AZ 85721, USA\\
$^{3}$Institute for Astronomy, University of Edinburgh, Royal Observatory, Blackford Hill, Edinburgh EH9 3HJ, UK\\
$^{4}$Institute of Astronomy, University of Cambridge, Madingley Road, Cambridge CB3 0HA, UK\\
$^{5}$Department of Physics and Astronomy, Western Washington University, 516 High St., Bellingham, WA 98225, USA\\
$^{6}$McWilliams Center for Cosmology, Department of Physics, Carnegie Mellon University, 5000 Forbes Avenue, Pittsburgh, PA 15213, USA\\
$^{7}$Department of Astronomy, The Ohio State University, 140 W. 18th Ave., Columbus, OH 43210, USA., and Center for Cosmology and AstroParticle Physics,\\ The Ohio State University, 191 W. Woodruff Ave., Columbus, OH 43210, USA.\\
$^{8}$Department of Statistics, Carnegie Mellon University, 5000 Forbes Avenue, Pittsburgh, PA 15213, USA\\
$^{9}$Department of Astronomy, University of Virginia, 530 McCormick Road, Charlottesville, VA 22904-4325, USA\\
$^{10}$Space Telescope Science Institute, 3700 San Martin Dr. Baltimore MD 21218\\
$^{11}$Department of Physics, Geology, and Engineering Technology, Northern Kentucky University, Highland Heights, KY 41099\\
$^{12}$Department of Physics \& Astronomy, Vanderbilt University, Nashville, TN 37235\\
$^{13}$Department of Physics \& Astronomy, Texas Christian University, Fort Worth, TX 76129, USA\\
$^{14}$Department of Physics, Montana State University, P.O. Box 173840, Bozeman, MT 59717-3840\\
$^{15}$NSF's National Optical-Infrared Astronomy Research Laboratory, 950 North Cherry Ave, Tucson, AZ 85719\\
$^{16}$Centro de Astronom{\'i}a (CITEVA), Universidad de Antofagasta, Avenida Angamos 601, Antofagasta 1270300, Chile\\
$^{17}$Center for Computational Astrophysics, Flatiron Institute, Simons Foundation, 162 Fifth Avenue, New York, NY 10010, USA\\
$^{18}$Departamento de F\'{i}sica, Facultad de Ciencias, Universidad de La Serena, Cisternas 1200, La Serena, Chile\\
$^{19}$Department of Physics, Salisbury University, Salisbury, MD 21801\\
}

\date{Accepted XXX. Received YYY; in original form ZZZ}

\pubyear{2020}

\begin{document}
\label{firstpage}
\pagerange{\pageref{firstpage}--\pageref{lastpage}}
\maketitle

\begin{abstract}
We use observations from the APOGEE survey to explore the relationship between stellar parameters and multiplicity. We combine high-resolution repeat spectroscopy for 41,363 dwarf and subgiant stars with abundance measurements from the APOGEE pipeline and distances and stellar parameters derived using \textit{Gaia} DR2 parallaxes from \cite{Sanders2018} to identify and characterise stellar multiples with periods below 30 years, corresponding to \drvm$\gtrsim$ 3 \kms, where \drvm\ is the maximum APOGEE-detected shift in the radial velocities. Chemical composition is responsible for most of the variation in the close binary fraction in our sample, with stellar parameters like mass and age playing a secondary role. In addition to the previously identified strong anti-correlation between the close binary fraction and \feh\, we find that high abundances of $\alpha$ elements also suppress multiplicity at most values of \feh\ sampled by APOGEE. The anti-correlation between $\alpha$ abundances and multiplicity is substantially steeper than that observed for Fe, suggesting C, O, and Si in the form of dust and ices dominate the opacity of primordial protostellar disks and their propensity for fragmentation via gravitational stability. Near \feh{} = 0 dex, the bias-corrected close binary fraction ($a<10$ au) decreases from $\approx$ 100 per cent at \alh{} = $-$0.2 dex to $\approx$ 15 per cent near \alh{} = 0.08 dex, with a suggestive turn-up to $\approx$20 per cent near \alh{} = 0.2. We conclude that the relationship between stellar multiplicity and chemical composition for sun-like dwarf stars in the field of the Milky Way is complex, and that this complexity should be accounted for in future studies of interacting binaries.
\end{abstract}




\section{Introduction}
\label{sec:intro}

The accurate characterisation of stellar multiplicity remains a key priority in stellar astrophysics. Interacting binaries, defined as those that are close enough to transfer mass and experience significant deviations from single stellar evolution, are responsible for a wide array of phenomena in time-domain astronomy. These include, but are not limited to, cataclysmic variables, novae, all Type Ia and many core-collapse supernovae, high- and low-mass X-ray binaries, and the majority of gravitational wave sources in the LIGO and LISA passbands \cite[for a review, see][]{Marco2017}. The formation rates of these sources in a variety of stellar populations are determined by the initial conditions for stellar multiplicity: the multiplicity fraction, and the distribution of periods, mass ratios, and eccentricities. It is now clear that these fundamental statistics of stellar multiplicity are strong functions of stellar properties like mass and composition, and that they are not independent of each other \citep[see][for reviews]{Duchene2013,Moe2017}. This realisation sets the stage for the challenging observational problem of identifying and characterising all the relevant correlations between stellar properties and multiplicity statistics in the field.

Fortunately, modern spectroscopic surveys are well suited to this task. The Apache Point Observatory Galactic Evolution Experiment 2 \citep[APOGEE-2,][]{Majewski2017}, one of the constituent surveys in the fourth instalment of the Sloan Digital Sky Survey \citep[SDSS-IV,][]{Gunn2006, Blanton2017}, collected multi-epoch data for 437,485 stars with its high-resolution (R$\sim$22,500) multiplexed infrared spectrograph as part of Data Release 16 (DR16) \citep{Wilson2019}. This constitutes the most comprehensive sample of the detailed compositions of Milky Way stars to date. The APOGEE Stellar Parameter and Chemical Abundances Pipeline \citep[ASPCAP,][]{Perez2016,Joensson2020} has measured reliable stellar parameters for each of these stars, including calibrated abundances of as many as 20 or more elements, and precise radial velocities \citep[RVs,][]{Nidever2015} for each individual visit spectra. Leveraging the time-domain component of the survey, \cite{Badenes2018} identified a strong anti-correlation between the multiplicity fraction at short periods and stellar metallicity in DR13 of APOGEE \citep{Albareti2017} - see also \cite{Grether2007, Raghavan2010, Gao2014, Gao2017, Yuan2015, El-Badry2018, El-Badry2018b, Pawlak_2019, Liu2019, Price-Whelan2020, Miglio2020}. Further analysis by \cite{Moe2019} established that the metal-poor (\feh$\sim-1$ dex) dwarfs observed by APOGEE are $\sim$4 times more likely to have short-period ($P\lesssim$30 yr, or $a\lesssim10$ AU) binary companions than the metal-rich (\feh$\sim$0.5 dex) dwarfs, and that this trend likely extends to the lower metallicities characteristic of halo stars. This anti-correlation has now been firmly established using large numbers of sparsely sampled RV curves \citep{Gao2014,Gao2017,Badenes2018,Price-Whelan2020}, smaller numbers of systems with known orbital periods \citep[from both complete orbital solutions and eclipses,][]{Moe2019}, and common-proper-motion binaries with projected separations measured by Gaia \citep{El-Badry2018}. This has profound implications for the rates of interacting binaries in the Universe \citep[e.g.][]{Paczynski1971,Iben1984,Suda2013,Mink2015,Marco2017,Breivik2019,Stanway2020} and for the physics of star formation and disk fragmentation \citep[e.g.,][]{Kratter2010a, Duchene2013, Moe2017, Moe2018a, Kounkel2019}. 

Here we continue to explore the relationship between stellar parameters and stellar multiplicity using public data from APOGEE, complemented by Gaia Data Release 2. In order to avoid the details of the interplay between stellar evolution and multiplicity described by \cite{Badenes2018}, we restrict our analysis to dwarf and subgiant stars. We examine a wide array of stellar parameters, paying special attention to the abundances of $\alpha$ elements. In Section~\ref{sec:samp}, we detail our sample selection and method to account for double-lined spectroscopic binaries (SB2s). Section~\ref{subsec:MCcorrec} describes our completeness corrections. In Section~\ref{subsec:triangle} we describe the broad view of the relationship between stellar multiplicity and stellar parameters in our sample. In Section~\ref{subsec:chemCBF}, we examine in more detail the impact of chemical composition on stellar multiplicity. We discuss our results in Section~\ref{sec:discuss} and summarise in Section~\ref{sec:concl}.

\section{Sample Selection}
\label{sec:samp}

\begin{figure*}
	\includegraphics[width=1.0\textwidth]{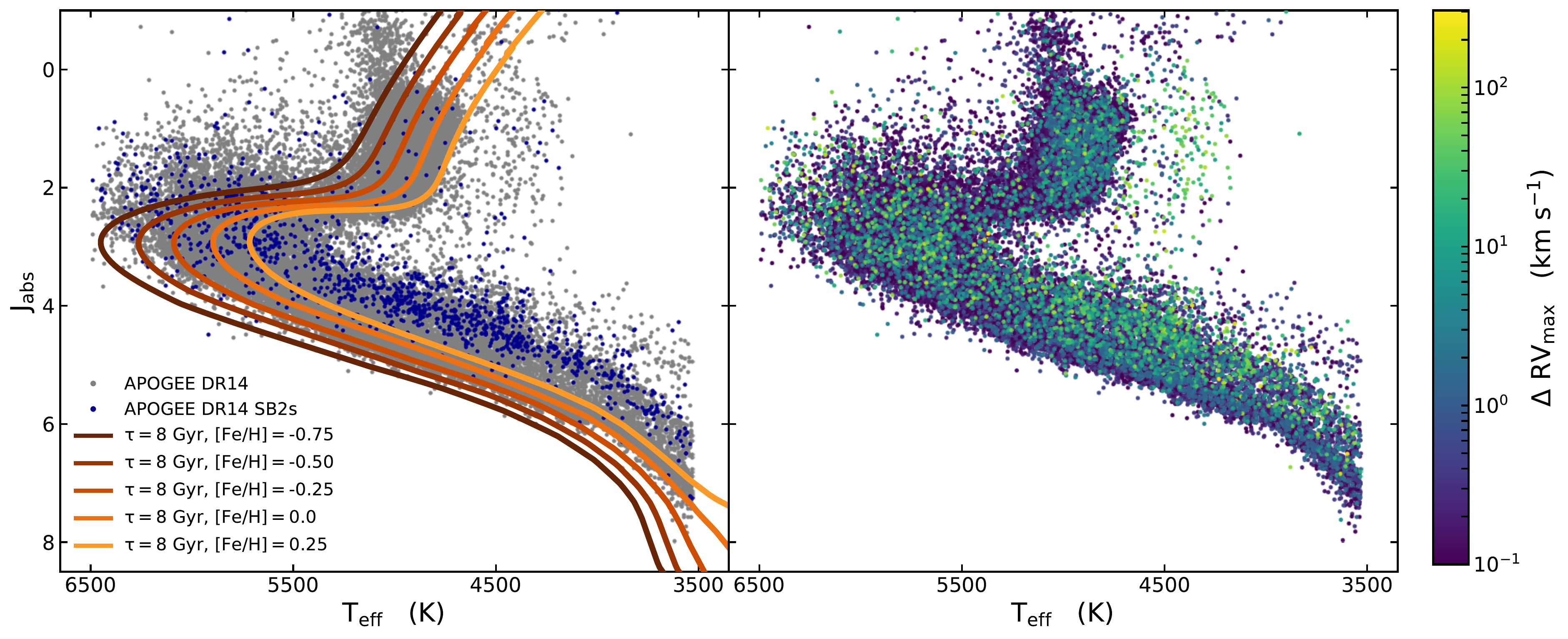}
    \caption{Left panel: An HR diagram for our sample with APOGEE DR14 uncalibrated \Teff{} and the absolute 2MASS $J$ magnitude, calculated using the \citet{Sanders2018} distance estimates. The grey points are for our main sample, and the dark blue are for objects identified as likely SB2s. The coloured lines are MIST isochrone tracks for $\tau =8$ Gyr and various metallicities. Right panel: The same HR diagram but with a colourbar on \drvm{}. Points with \drvm{}$\geq1$ \kms{} are plotted on top for clarity.}
    \label{fig:HRdiag}
\end{figure*}

The DR14 version of the APOGEE \verb+allStar+ file contains spectral parameters for 277,371 entries \citep{Abolfathi2018, Holtzman2018, Joensson2018}. We first note that there are only 258,475 unique APOGEE IDs amongst these 277,371 entries. The duplicate entries are a result of a star being observed in different fibre plugplates with different field centres, which are not automatically combined by the pipeline. Each \verb+allStar+ entry corresponds to a combined spectrum and its measured stellar parameters, and is uniquely described by an APOGEE ID and a field location ID.

From APOGEE DR14, we removed stars with the STAR\_BAD flag set in the ASPCAP bitmask \citep{Holtzman_2015} and those targeted as telluric calibrators \citep[bit 9 in both the apogee\_target2 and apogee2\_target2 masks;][]{Zasowski2013, Zasowski2017}. Star cluster members (bit 9 in apogee\_target1 and apogee2\_target1 and bit 10 in apogee\_target2 and apogee2\_target2) and commissioning stars \citep[bit 1 in STARFLAG,][]{Holtzman_2015} were removed as well. Finally, we required acceptable ($\neq-9999$, APOGEE's default for a bad value) uncalibrated effective temperatures (\Teff{}) and surface gravities (\logg{}) to maximise our ability to distinguish dwarfs from giants in DR14. As noted in \citet{Holtzman2018}, dwarfs in APOGEE DR14 do not have calibrated \logg{} values, so we do not make cuts on the calibrated parameters. In order to estimate the dereddened $JHK_{s}$ magnitudes, we used the value of $A_{\text{K}}$ adopted for targeting purposes \citep[AK\_TARG,][]{Zasowski2013, Zasowski2017}. 

For each APOGEE ID/location ID combination, we identified the individual visits from the \verb+allVisit+ file that were included in its combined APOGEE spectrum \citep[the VISITS\_PK indices,][]{Holtzman_2015, Nidever2015}. We imposed an additional quality cut, requiring two or more of these visits to have a S/N$\geq40$. If a star had duplicate APOGEE IDs, all of the acceptable visit RVs from its various plugplate fields were concatenated. This meant that objects with at least one acceptable visit in two or more fields could be included. For these stars, we averaged any duplicate stellar parameters with valid values from the pipeline.

Both the APOGEE data reduction pipeline \citep[][]{Nidever2015} and ASPCAP \citep[][]{Perez2016} assume that each source can be modelled by a single stellar spectrum. Stellar companions within the range of mass ratios $q=M_{2}/M_{1}$ that can make a significant contribution to the observed flux (double-lined spectroscopic binaries or SB2s) can therefore introduce biases in the spectral fits; see \citet{El-Badry2018a} for a discussion. To identify these stars, we examined the APOGEE cross-correlation functions (CCFs), following the procedure described in \citet{Kounkel2019}. Two approaches were considered: using CCFs that APOGEE provides natively in its data releases, and recalculating the CCFs by cross-matching the spectra with the best-fit PHOENIX synthetic spectrum \citep{Husser2013}, using the reported RV\_TEFF and RV\_LOGG parameters. In most cases, the deconvolution of multiple components from the CCFs occurred in the same sources, with comparable RVs. In this way, we identified 3656 likely SB2s within APOGEE DR14, of which 1512 were in our quality-cut sample. From the CCFs for these stars, we determined the RV of the highest peak at each epoch and used this as a more reliable estimate for the RV of the photometric primary. After applying our quality cuts, we were left with 1495 likely SB2s, which we kept in our sample with spectral parameters from APOGEE/ASPCAP and RVs from our CCF analysis. Details about the downloadable tables of these SB2s are available in Appendix~\ref{appx:sb2s}.

In a final step, we restricted our sample to \loggunits{} $\geq3.25$, \feh{} $\geq-1.0$ dex. This simple cut in \logg{} will not purely select dwarfs, but it is sufficient for our purposes in eliminating most stars on the red giant branch. We also imposed an additional requirement of acceptable values ($\neq-9999$) for \alfe{}, \alh{}, \oh{}, \mgh{}, and \sih{}. This left us with 41,363 unique APOGEE targets, 1278 of which were identified as SB2s, and 3896 (131 SB2s, 3765 non-SB2s) had duplicate entries and so their stellar parameters were averaged. The fraction of SB2s in this sample is 1278/41,363 = 3.1 $\pm$ 0.1 per cent, consistent with the 2.8 $\pm$ 0.2 per cent value measured in young stellar objects by \cite{Kounkel2019}.

Unlike \citet{Kounkel2019}, which focused primarily on the young stellar objects, most of the sources deconvolved as SB2s in this work are main sequence stars, and their CCFs are not affected by variability due to star spots. Therefore, it is possible to reliably include sources with quality flag 3 in addition to 4 in the list of likely SB2s (see Table 5 and Section 4.1 in \cite{Kounkel2019} for an explanation of these flags). Thus, we caution against blindly comparing these fractions. \cite{El-Badry2018b} used a more sophisticated method based on \emph{The Payne} \citep{Ting2019}, to identify SB2s from RV shifts among dwarf stars in APOGEE DR12. Their measured SB2 fraction from this method is 663/20,142 = 3.3 $\pm$ 0.1 per cent, which is consistent with our results. These authors also found SB2s by making multi-component spectral fits, and found a higher SB2 fraction of 2645/20,142 = 13.1 $\pm$ 0.2 per cent. However, many of the systems identified by this method had small or negligible RV shifts and therefore this higher SB2 fraction is hard to compare with what we measure in our RV-selected sample.

We cross-matched our final sample of APOGEE targets with the catalogue from \citet{Sanders2018}, who calculated Bayesian posteriors on distance $d$, mass $M$, and age $\tau$, by fitting PARSEC isochrones to a combination of Gaia DR2 parallaxes, broadband photometry, and the spectral parameters derived by ASPCAP. \citet{Sanders2018} give non-NAN values of $d$, $M$ and $\tau$ for the vast majority (41,014, or 99 per cent) of the stars in our sample. We use these distance estimates to plot absolute 2MASS magnitudes $J_{\text{abs}}$ vs. uncalibrated APOGEE \Teff{} in Fig.~\ref{fig:HRdiag}. The left panel shows the bulk of our sample in grey with the SB2s over-plotted in dark blue. Isochrones from the MESA Isochrone and Stellar Tracks Collaboration \citep[MIST;][]{Paxton2011,Paxton2013,Paxton2015,Dotter_2016,Choi2016} are shown for $\tau=8$ Gyr and a range of representative metallicities. According to \cite{Sanders2018}, the age distribution in our sample peaks around 8 Gyr, which is in good agreement with the main sequence turn-off point shown in Fig.~\ref{fig:HRdiag}. The majority of the SB2s lie above the single star isochrone tracks, as expected for systems with a measurable flux contribution from both components. The right panel shows the same HR diagram coloured by the maximum shift in the RVs, \drvm{} \citep[see][]{Badenes2012,Maoz2012,Badenes2018,Moe2019}, with stars that have \drvm{}$\geq1$ \kms{} plotted on top for clarity. Here too we find a significant excess of objects with large RV variability to have locations above the single-star isochrones.

\section{Results}
\label{sec:results}

\subsection{Stellar multiplicity, \drvm\ distributions, and completeness corrections}
\label{subsec:MCcorrec}

\begin{figure}
	\includegraphics[width=\columnwidth]{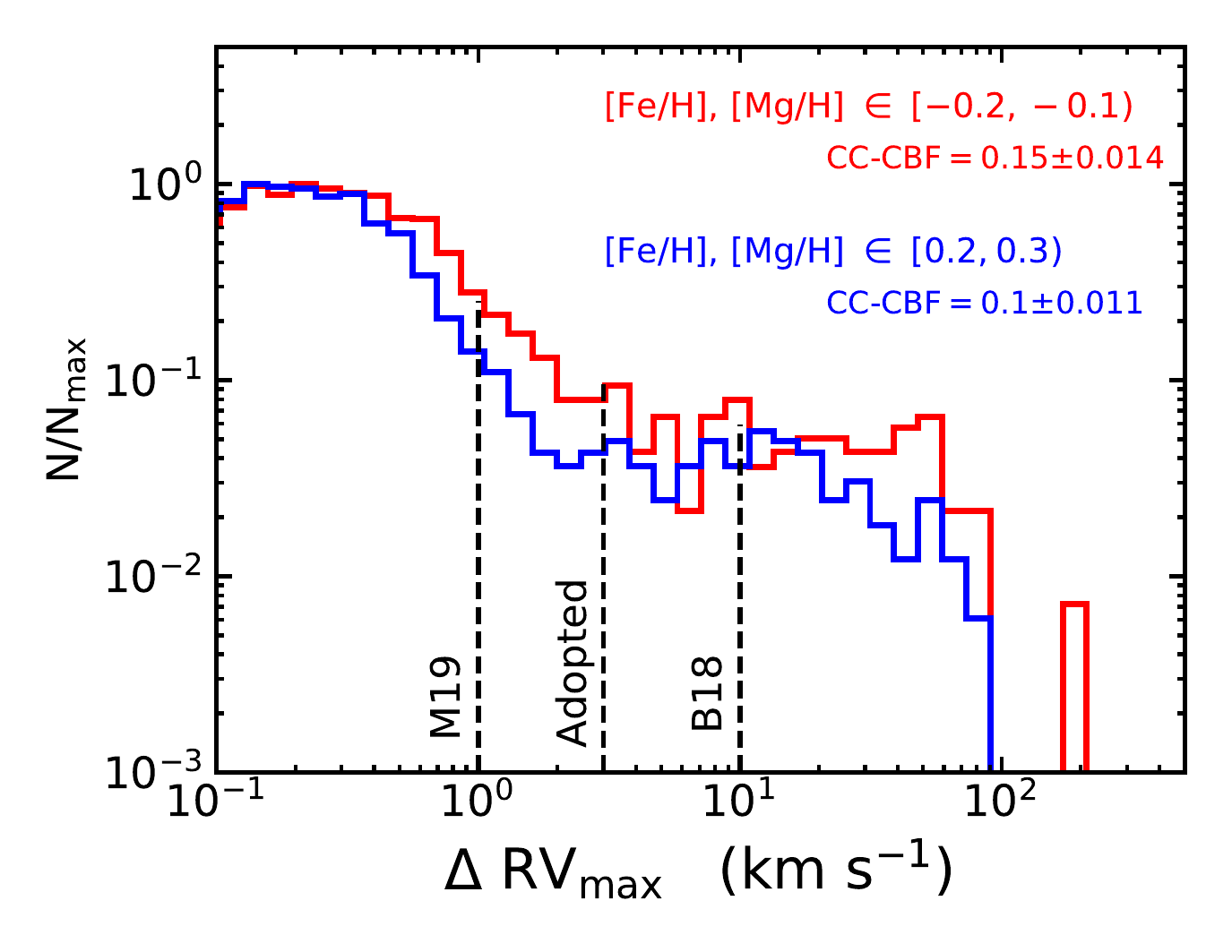}
    \caption{Normalised distributions of \drvm{} for two boxes with $N\sim2000$ in \feh-\mgh{} space from Fig.~\ref{fig:triangle}. The \drvm{} thresholds from \citet{Moe2019}, \citet{Badenes2018}, and the present work are shown as dashed lines.}
    \label{fig:hists}
\end{figure}

Following \cite{Badenes2012}, \cite{Maoz2012}, and \cite{Badenes2018}, we use \drvm\ as a figure of merit to evaluate the sparsely sampled RV curves from APOGEE. Most (42.9 per cent) of the stars in our sample have 3 visits, with 36.4 per cent having 2 and the rest having 4 or more. While this is not enough to define a full orbital solution for most stars \citep[see][for discussions]{Price-Whelan2018, Price-Whelan2020}, values of \drvm\ above a certain threshold can securely identify large numbers of short-period binaries. In Fig.~\ref{fig:hists}, we show the distribution of \drvm\ in two groups of $N\sim2000$ stars with constant \feh{} and \mgh{}. This example illustrates the two main features of \drvm\ distributions derived from high quality data: a core of low \drvm\ values dominated by measurement errors and an extended tail of high \drvm\ values dominated by stars with companions in short-period orbits, clearly defined and cleanly separated from the core. We refer the reader to the discussions in \cite{Badenes2018} for the role of measurement errors, metallicity, and RV jitter in the APOGEE \drvm\ distributions. Here we focus on two closely related issues: the completeness corrections and the threshold value of \drvm\ to single out multiple systems.

We estimate completeness corrections on these \drvm\ distributions with a Monte Carlo sampler similar to that used by \cite{Moe2019}. Our sampler simulates a population of $N$ systems, with the fraction of systems in binaries determined by a free parameter called the multiplicity fraction $f_{m}$. Each system is assigned a visit history (number of visits and time lags between visits) from a random star in our APOGEE DR14 dwarf/subgiant sample. For each simulated binary, we draw the main orbital parameters (period and eccentricity) from the observational distributions measured for field solar-type binaries (period, \citealt{Raghavan2010}; eccentricity, \citealt{Moe2017}), select a random orbital inclination and initial phase, and generate RVs by sampling the projected orbit with the visit history, adding RV errors from a user-specified distribution. For each simulated single star, we set all RVs to zero and add errors from the same distribution. The code is described in more detail in \cite{Badenes2018} -- here we list the specific choices made for the present work. We simulate $N=50,000$ stars with $f_{m}=0.5$. Each star is assigned \loggunits$=4.25$, the median value for our sample, which corresponds to a critical Roche Lobe Overflow period of $P_{\text{crit}}=0.49$ days in a 1 \msun\ binary with $q=1$. The primary mass $M$ is randomly drawn from the distribution of \cite{Sanders2018} mass estimates for our sample (shown in the second diagonal panel of Fig.~\ref{fig:triangle}). For the mass ratio $q$, we assume a flat distribution with a twin excess fraction of 25 per cent for systems with $0.95\leq q \leq1.0$ \citep{Moe2017}. The RV errors are drawn from a Student's t distribution (scipy.stats.t) with degrees of freedom 3.5, location 0 and scale 0.25. Appendix~\ref{appx:MC} discusses these choices and their effects on the completeness corrections in more detail.

\begin{figure*}
	\includegraphics[width=\textwidth]{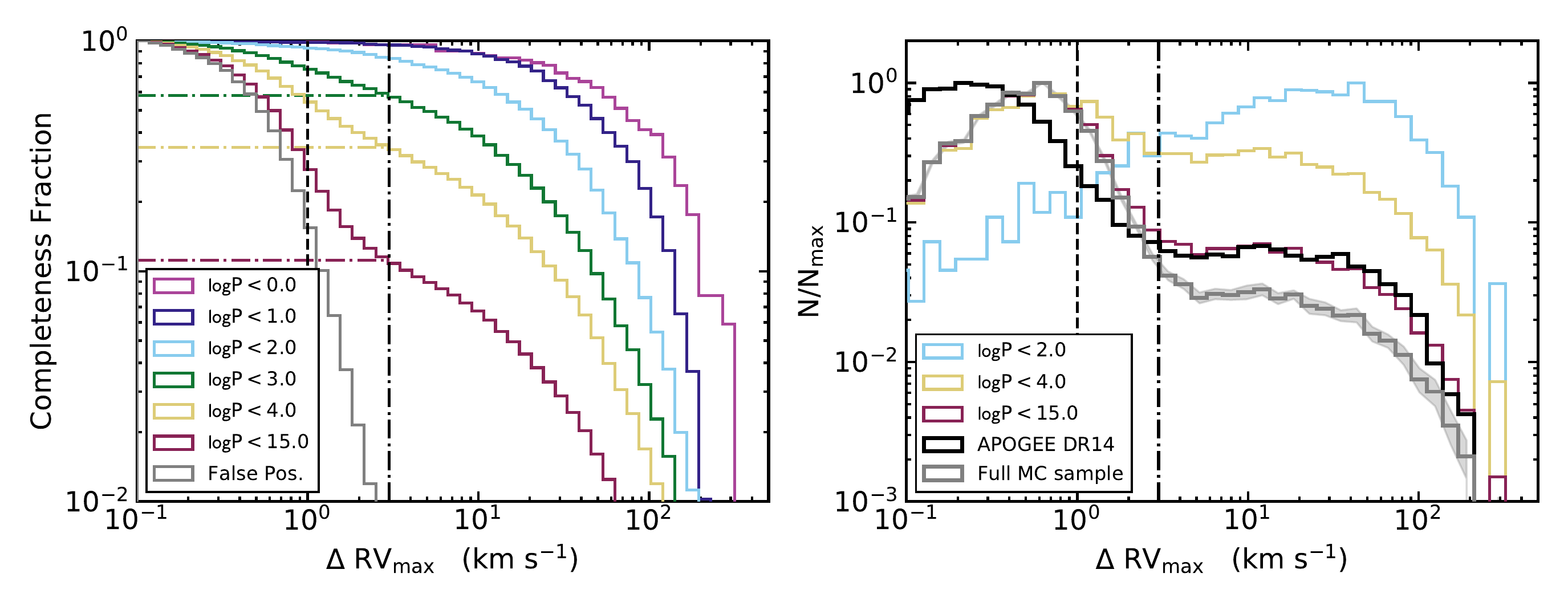}
    \caption{Left panel: Completeness fraction versus \drvm{} for Monte Carlo generated samples at various period limits, using the APOGEE DR14 time lags from our sample. The horizontal dot-dashed lines are the completeness fractions for the relevant $\log(P/\text{days})$ samples given a threshold \drvm{}$\geq3$ \kms{}. The vertical dashed line is at \drvm{}$=1$ \kms{} for comparison. The grey histogram shows the completeness fraction for false positives--systems that are not binaries but show some RV variation due to simulated RV error. Right panel: normalised \drvm{} distributions for MC data. The fainter lines are for several $\log(P/\text{days})$ samples, colour-coded as in the left panel. The grey histogram is for the full MC sample (binaries and non-binaries included), with shading indicating 1$\sigma$ intervals from bootstrapping the sample ($N_{\text{boots}}=25$). The black histogram is for our APOGEE DR14 sample.}
    \label{fig:compfracs}
\end{figure*}

\begin{table*}
	\centering
	\caption{Completeness fractions for selected $\log(P/\text{days})$ and \drvm{} thresholds.}
	\label{tab:compfrac}
	\begin{tabular}{lccc}
		\hline
		$\log(P/\text{days})$ threshold & \drvm{}$\geq1$ \kms{} & \drvm{}$\geq3$ \kms{} & \drvm{}$\geq10$ \kms{}\\
		\hline
		$\log{P}\leq0.0$ & 1.00 & 0.96 & 0.88 \\
		$\log{P}\leq2.0$ & 0.93 & 0.84 & 0.66 \\
		\hline
		$\log{P}\leq4.0$, this work & 0.55 & 0.34 & 0.21 \\
		$\log{P}\leq4.0$, from \cite{Moe2019} & 0.57 & 0.40 & 0.24 \\
		\hline
		$\log{P}\leq15.0$ & 0.29 & 0.11 & 0.07 \\
		False Positives & 0.17 & 0.005 & 0.00 \\

		\hline
	\end{tabular}
\end{table*}

In Table~\ref{tab:compfrac} and the left panel of Fig.~\ref{fig:compfracs}, we show the cumulative fraction of systems with \drvm\ above a given value in several period ranges in our Monte Carlo simulation. Assuming the underlying period and eccentricity distributions are not too different from the assumed ones, the completeness correction that needs to be applied to recover the total number of binaries with periods below a certain value is the inverse of these cumulative fractions. Our results are consistent with those of \cite{Moe2019} (included in Table~\ref{tab:compfrac}), who applied a similar approach to a sample of APOGEE DR13 dwarfs. The grey and red histograms in Fig.~\ref{fig:compfracs} show the cumulative fractions for all non-binary and binary systems, respectively. The false positive rate for binaries in a given period range at a given value of \drvm\ is the ratio between the relevant cumulative fraction and the grey histogram at that value of \drvm.

These curves inform our choice of \drvm\ threshold value. A conservative value like the 10 \kms\ chosen by \cite{Badenes2018} is virtually free of false positives, but results in low detection efficiencies and correspondingly large completeness corrections, which can lead to issues when dealing with small samples of systems with a specific set of stellar parameters. For the dwarf and subgiant stars that we examine here, which have low RV jitter \citep{Hekker_2008} and relatively narrow \drvm\ distribution cores \citep{Badenes2018}, we propose a more reasonable value of 3 \kms{}. Using the uncertainties reported by the APOGEE data reduction pipeline, the median RV uncertainty for our sample is $\sigma_{RV}$ $\sim$ 0.04 \kms, though these uncertainties are almost certainly underestimated \citep[see discussions in][and sources within]{Holtzman2018, Badenes2018}. We can instead consider a more reasonable value of $\sigma_{RV}\sim$ 0.2 \kms, obtained from roughly fitting the observed \drvm{} distribution core to those simulated by our MC with Gaussian error distributions with mean of 0 and varying spreads \citep[similar to APOGEE DR13, see Fig.~9 of][]{Badenes2018}. Regardless, our threshold remains far larger than what can be explained with typical RV uncertainties alone. This threshold yields a detection efficiency of $\approx 34$ per cent for systems with $\log(P/\text{days})\leq4.0$ and $\approx 84$ per cent for $\log(P/\text{days})\leq2.0$, with an overall false positive rate of $\approx 0.1$ per cent. Compared to \cite{Moe2019}, who chose a threshold value of \drvm{}$\geq1$ \kms{}, we expect a false positive rate about $30$x lower, with only a modest loss of $\approx20$ per cent in detection efficiency.

In the context of our APOGEE sample, completeness corrections for systems with $\log(P/\text{days})>4$ ($a>10$ AU) are unwarranted for several reasons. These long-period binaries will rarely produce detectable RV variability in APOGEE, and are often difficult to characterise using sparsely sampled RV curves. Moreover, the anti-correlation between stellar multiplicity and \feh\ weakens beyond $a>50$ AU and disappears beyond $a>200$ AU \citep{Moe2019,El-Badry2018}, and this might apply to other stellar parameters. Therefore, in the remainder of this work we will quote completeness-corrected binary fractions for systems with $\log(P/\text{days})\leq4$, which we identify as `close binaries'. For reference, a 1 \msun\ star of solar composition at the tip of the red giant branch has a critical period for Roche Lobe overflow of $\log(P/\text{days}) \sim 2.8$.

In the right panel of Fig.~\ref{fig:compfracs} we compare the simulated \drvm{} distribution from our MC run to the observed distribution in the APOGEE sample. We estimate $1\sigma$ intervals on the simulated distribution (shown as the grey shading) by bootstrapping the sample with $N_{\text{boot}}=25$, $N_{\text{sys}}=40,000$. We also show the \drvm{} distributions in three different subsets of simulated systems: those with $\log(P/\text{days})<15$ (all binaries), $\log(P/\text{days})<4$ (all close binaries), and $\log(P/\text{days})<2$. We do not attempt to provide an accurate match to the observed \drvm{} distribution, as this would require a complete characterisation of the correlations between stellar properties, multiplicity, and RV errors, but we note that the shape and extent of the tail in our simulation is very similar to what we see in the APOGEE sample. We also note that our choice of RV error distribution is conservative, as shown by the comparison between the simulated and observed core shapes. 

Binaries in general, and twins in particular, can be detected further away than single stars in magnitude-limited samples due to Malmquist bias (see Fig.~\ref{fig:HRdiag}). Conversely, it is more difficult to detect RV variability of twin SB2s if their absorption features are significantly blended \citep[but see][for an alternative approach]{El-Badry2018b}. In their analysis, \cite{Moe2019} estimated that these two effects bias the close binary fraction measured by APOGEE by $\approx$ 30 per cent in opposite directions and therefore approximately cancel each other. However, they relied solely on the APOGEE pipeline RV measurements, while we applied a CCF method to identify SB2s and more accurately measure their RVs. Our Malmquist bias in favour of detecting twin binaries should therefore be slightly greater than our inefficiency in the detection of RV variability in SB2s. We compensate for this by reducing our completeness-corrected close binary fractions by 10 per cent to make the reported values more representative of volume-limited samples. This results in an estimated detection efficiency of 0.38 for \drvm{}$\geq3$ \kms{} and $\log(P/\text{days})\leq4$, which we adopt for the remainder of this work. Using this completeness correction, the close binary fractions we recover from the \drvm\ distributions shown in Fig.~\ref{fig:hists} are $0.15\pm0.014$ and $0.1\pm0.011$.

\begin{figure*}
	\includegraphics[width=\textwidth]{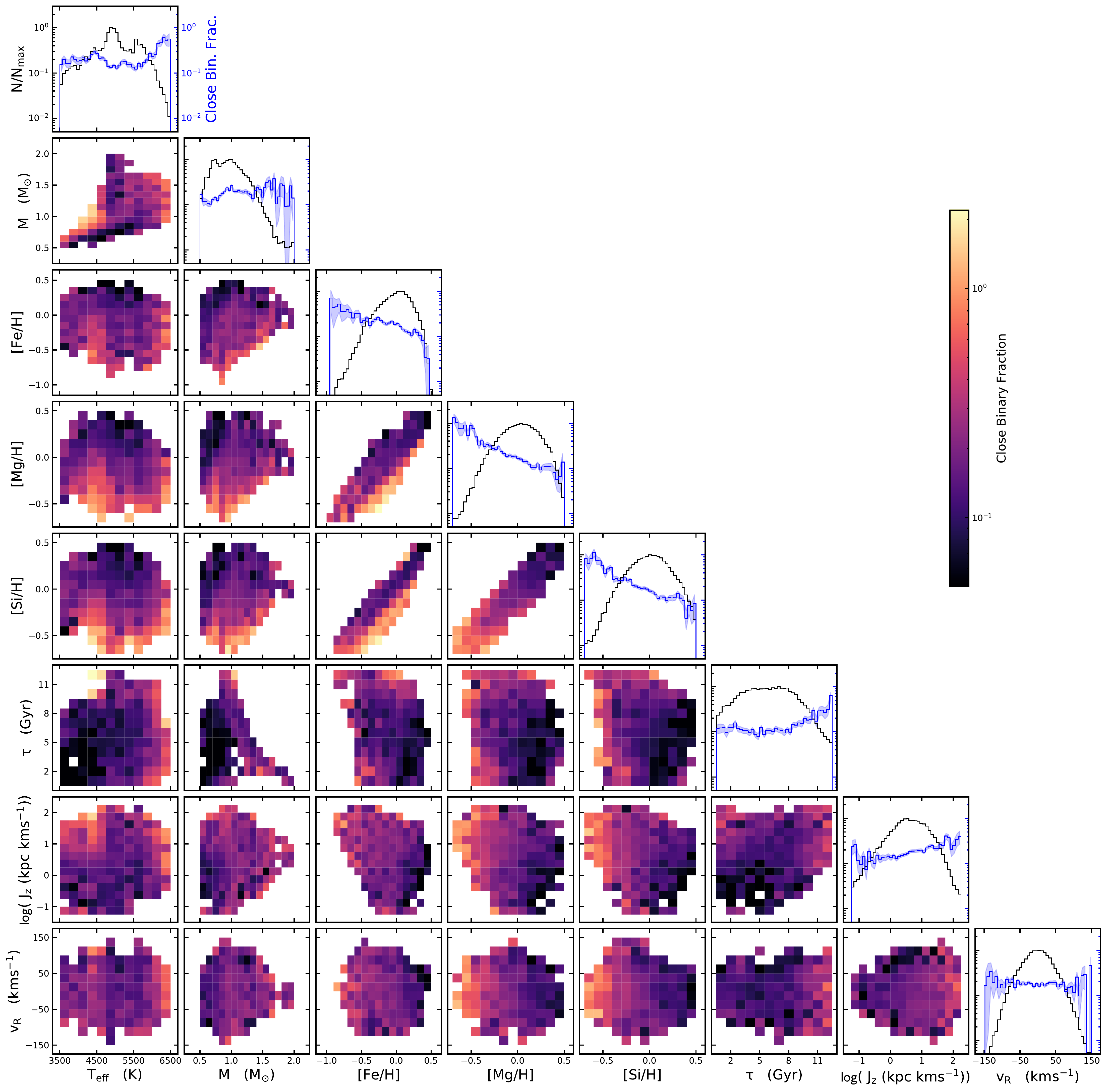}
    \caption{Two-dimensional histograms showing the completeness-corrected close binary fraction as a function of many of the parameters considered in this work. Along the diagonal, the black lines are the normalised histograms of each parameter, and shown in blue is the completeness-corrected close binary fraction as a function of that parameter alone. The blue shaded region shows the uncertainties on the completeness-corrected close binary fraction (equation~\ref{eq:PN}).}
    \label{fig:triangle}
\end{figure*}

\subsection{The Impact of Stellar Parameters on the Close Binary Fraction}
\label{subsec:triangle}

We are now in a position to examine the impact of stellar parameters on the completeness-corrected close binary fractions measured in our sample of APOGEE dwarfs and subgiants. To do this, we choose a few representative parameters among those measured by APOGEE and \cite{Sanders2018}: \Teff, $M$, \feh, \mgh, \sih, $\tau$, the vertical action $J_{z}$, and the galactocentric radial velocity $v_{R}$. The vertical action is defined as 
\begin{equation}
    J_{z} = \frac{1}{2\pi}\oint dz v_{z}
\end{equation}
where $z$ and $v_{z}$ are the position and galactocentric vertical velocity for the star along its orbit. As an indicator of a star's vertical displacement, $J_{z}$ is unaffected by orbital phase as compared to $z$ or $v_{z}$, and it is a tracer of the birth location of stars in the Milky Way disk that is more robust to radial migration than galactocentric radius \citep{Vera-Ciro2014}.

\begin{table*}
	\centering
	\caption{Fit parameters and the number of $\sigma$ (estimated via bootstrapping) for the parameter to be consistent with 0.}
	\label{tab:fitparams}
	\begin{tabular}{|c|cc|cccc|}
	     & \multicolumn{2}{|c|}{$\log{f_{m}}=b+aX$} & \multicolumn{3}{|c|}{$\log{f_{m}}=c+bX+aX^{2}$} & \\
		\hline
		 & $b$ & $a$ & $c$ & $b$ & $a$ & $\chi^{2}_{\text{lin}} /\chi^{2}_{\text{quad}}$ \\
		\hline
		 
		\multirow{2}{*}{\Teff\  (K)} & -1.595  & $1.9e$-7  & 3.318   & -0.002   & $1.9e$-7 & \multirow{2}{*}{2.7}   \\
	                           &  & \textcolor{violet}{$2.7\sigma$} & & \textcolor{blue}{$5.4\sigma$}   & \textcolor{blue}{$5.8\sigma$} & \\
	 	
		\hline
		\multirow{2}{*}{$M$  (\msun)}   & -0.815  & 0.11      & -1.329  & 0.993    & -0.34   & \multirow{2}{*}{1.5}    \\
                               &  & \textcolor{violet}{$2.1\sigma$} & & \textcolor{violet}{$3.1\sigma$} & \textcolor{violet}{$2.5\sigma$} & \\
		
		\hline
		\multirow{2}{*}{\feh\  (dex)}  & -0.787  & -0.595    & -0.782  & -0.436   & 0.196   & \multirow{2}{*}{0.98}   \\
                               &  & \textcolor{blue}{$8.6\sigma$}  & & \textcolor{violet}{$4.7\sigma$}  & $1.2\sigma$ & \\
		
		\hline 
		\multirow{2}{*}{\mgh\  (dex)}  & -0.806  & -1.32     & -0.752  & -0.851   & 0.627   & \multirow{2}{*}{3.5}    \\
                               &  & \textcolor{blue}{$13.0\sigma$} & & \textcolor{blue}{$6.6\sigma$}    & \textcolor{violet}{$2.7\sigma$} & \\
		
		\hline 
		\multirow{2}{*}{\sih\  (dex)} & -0.77    & -1.15     & -0.755  & -1.011   & 0.187   & \multirow{2}{*}{1.3}     \\
                               &  & \textcolor{blue}{$9.1\sigma$}  & & \textcolor{violet}{$3.6\sigma$}  & $0.5\sigma$ & \\
		
		\hline
		\multirow{2}{*}{$\tau$  (Gyr)}& -1.355  & 0.077     & -0.696  & -0.127   & 0.013   & \multirow{2}{*}{4.7}     \\
                               &  & \textcolor{violet}{$2.8\sigma$} & & \textcolor{violet}{$3.5\sigma$} & \textcolor{violet}{$4.4\sigma$} & \\
		
		\hline
		\multirow{2}{*}{$\log(J_{z}/\text{kpc \kms} )$} & -0.76  & 0.142 & -0.783 & 0.045  & 0.062   & \multirow{2}{*}{1.4}     \\
                               &  & \textcolor{blue}{$5.0\sigma$}   & & $1.1\sigma$ & \textcolor{violet}{$2.7\sigma$} & \\
		
		\hline
		\multirow{2}{*}{$v_{R}$  (\kms)}& -0.662 & $1.0e$-5  & -0.757  & $1.0e$-4 & $1.0e$-5 & \multirow{2}{*}{1.1}   \\
                               &  & $0.1\sigma$                     & & $0.4\sigma$ & \textcolor{violet}{$2.4\sigma$} & \\
								
		\hline     
	\end{tabular}
\end{table*}

Several of these parameters are precisely determined by APOGEE (\Teff, chemistry), whereas others represent fundamental stellar properties ($M$, $\tau$) or are related to galactic dynamics that may prove interesting ($J_{z}$, $v_{R}$). Of course many of these parameters, like $\tau$ and \feh, have substantial internal correlations that cannot be properly examined without a multivariate analysis. Moreover, we are restricted to the parameter ranges covered by APOGEE, which are very broad for some parameters like \feh, but quite narrow for others that are of high interest for stellar multiplicity, like $M$. Finally, not all these parameters are equally well constrained by the observations. Stellar ages, for example, are notoriously hard to estimate without asteroseismic data \citep[e.g., see][]{Ness2016,Pinsonneault2018}. We also note that both ASPCAP and \cite{Sanders2018} \textit{assume} single star models, which can introduce biases in some parameters \citep[see][for a discussion]{El-Badry2018a}.

With all these caveats in mind, we present our view of the impact of stellar parameters on the close binary fraction in Fig.~\ref{fig:triangle}. This triangle plot shows the completeness-corrected close binary fraction as a two-dimensional histogram mapped on each pairwise combination of parameters. The one-dimensional terminal plots show the full distribution of each parameter in the APOGEE sample (black histograms) and the completeness-corrected close binary fraction as a function of that parameter alone (blue histograms with shaded error bars). We required a minimum of ten objects per bin in order to extend our measurements through the sample's full range of parameter space. Uncertainties are not shown in the 2D histograms, but they scale as $\sigma_{f}/c$, where $c=0.38$ is the completeness-correction discussed in Section~\ref{subsec:MCcorrec}, and $\sigma_{f}$ is the uncertainty from the binomial process on each measurement,
\begin{equation}
    \sigma_{f} = \sqrt{\frac{f(1-f)}{N}}    \label{eq:PN}
\end{equation}
where $f$ is the fraction of systems with \drvm{}$\geq3$ \kms{}, and $N$ is the total number of systems in that bin. Measurements made with small-$N$ samples will be noisy due to the $\sqrt{1/N}$ factor, but the RV variable fraction $f$ also introduces a $\sqrt{f(1-f)}$ factor. For a bin with $N=10$, we can consider two cases: (1) $f=0.2$ and (2) $f=0.8$. In both instances, the binomial process uncertainty is $\sigma_{f}/c=0.33$. The completeness-corrected close binary fractions are (1) $f_{m}=0.52\pm0.33$ and (2) $f_{m}=2.09\pm0.33$, showing that it is possible to measure variations in the close binary fraction even in bins with $N$ as small as 10.

Note also that our completeness-correction can result in close binary fractions that are in excess of 100 per cent, and we indeed see bins with values of $f_{m}\sim$2.0 in Fig.~\ref{fig:triangle}. We assumed the same period distribution for the entire simulated sample, and this assumption is most likely not valid across our diverse APOGEE sample. From the ASAS-SN Catalogue of Variable Stars, \cite{Jayasinghe2020} found that metal-poor eclipsing binaries were skewed towards shorter periods than metal-rich systems at fixed temperature. A shift towards shorter periods for metal-poor stars results in an over-correction from the completeness estimate, leading to our excessively large close binary fractions. Future studies of the period distribution as a function of chemistry and metallicity will be useful for addressing this issue.

The salient features of Fig.~\ref{fig:triangle} can be summarised as follows:

\begin{enumerate}
    \item The parameters related to chemical composition (\feh, \mgh, and \sih) emerge as the dominant drivers of stellar multiplicity in our sample. The completeness-corrected close binary fractions as a function of these parameters (blue 1D histograms in the diagonal panels) show clear monotonic downward trends, with dynamic ranges in excess of an order of magnitude, that are distinctly larger than for any other parameters. The gradients due to this downward trend are the most striking feature in all the 2D histograms that include chemical composition parameters. While the trends are uniform and monotonic in the 1D histograms, the 2D histograms reveal a great deal of complexity in the relationship between stellar multiplicity and chemical composition, which we examine in further detail in Section~\ref{subsec:chemCBF}.
    
     \item Even though stellar mass (and by proxy, \Teff) is known to have a strong effect on the close binary fraction of field dwarfs \citep[][]{Lada_2006,Duchene2013,Moe2017}, this is not clear in the APOGEE sample. The close binary fraction ($\log(P/\text{days})\lesssim4$) for Solar-mass stars scales as $M^{0.5}$ \citep{Moe2017}, so we expect the close binary fraction to increase by a factor of 2 across the sample's mass range. From the 1D histogram, we observe the close binary fraction increasing by a factor of $\sim1.5$. However, the close binary fraction measurements in the high-$M$ bins are noisy, and the mass estimates themselves are poorly constrained compared to APOGEE \Teff, so our measurement alone cannot be considered to be at odds with previous work. We do detect a noticeably higher close binary fraction for the hottest (\Teff $\gtrsim6000$ K) stars, which \cite{Price-Whelan2020} also found in a sample of binaries in APOGEE DR16. While this spike may be due to larger primary masses, the correlation at lower temperatures seem weaker. This might be due to the overlap between dwarfs and subgiants below 6000 K (apparent in Fig.\ref{fig:HRdiag}). In the \Teff-$M$ 2D histogram, there appears to be a region of increased binaries around 4000 K and 1 \msun, though this is more likely to be a result of erroneous mass estimates, given the temperature and mass values.
    
	\item Stellar age shows a modest upward trend, though this is hard to interpret. Stellar ages are poorly constrained in general, and age estimates for SB2s are particularly prone to errors: SB2 systems may be mis-classified as overly young (100s Myr) or overly old ($>10$ Gyr), because stars that are offset from the MS, like the high-\drvm{} objects in Fig.~\ref{fig:HRdiag}, may be classified along stellar pre-MS or post-MS tracks. This could account for the apparent increase in multiplicity fraction for $\tau>8$ Gyr. There is also a well established (though complex) correlation between \alfe{}, \feh{}, and age \citep[][and sources within]{Mackereth2017,Mackereth2019}, which is often used in studies of galactic dynamical evolution. A more complete treatment of these correlations is required before we can comment on any trends between age and the close binary fraction.
	
	\item  The 1D histogram for $\log{J_{z}}$ shows a significant correlation with the completeness-corrected close binary fraction, but this could simply be due to the fact that the outer disk is more metal-poor \citep[e.g., in APOGEE][and sources within]{Hayden2015,Weinberg2019}.
    
    \item The galactocentric radial velocity shows the flattest distribution of the parameters studied here. In the 1D histogram, the bins at either edge in parameter space appear to have an increased binary fraction, but they are consistent with a flat distribution given their large uncertainties.
\end{enumerate}

To quantify the impact of each parameter on the multiplicity fraction, we fit linear and quadratic functions to each of the blue histograms along the diagonals of Fig.~\ref{fig:triangle}. The best fit parameters and the ratio between the $\chi^{2}$ are given in Table~\ref{tab:fitparams}. None of the distributions are necessarily expected to follow a linear or quadratic function, but these are simple, easily-fit functions that provide an estimate of the slopes of the distributions. We then bootstrapped ($N_{\text{boot}}=500$) the fits to estimate uncertainties on the fit parameters. We can then calculate the number of $\sigma$ required for the first and second derivatives to be consistent with zero. These values are listed in the second row for each parameter in Table~\ref{tab:fitparams}, with significant values ($n>2$) in purple and highly significant values ($n>5$) in blue. From these values, we conclude that the chemical composition parameters show the most significant correlations with close binary fraction in our sample, though there are also clear trends with stellar age, mass, \Teff, and vertical action. We recover the strong anti-correlation between \feh{} and the completeness-corrected close binary fraction previously reported by various authors, and identify for the first time a similar effect in both sign and strength for $\alpha$-process elements Mg and Si. Characterising these correlations is the subject of the remainder of this paper.

\subsection{Chemical Composition and the Close Binary Fraction}
\label{subsec:chemCBF}

\begin{figure}
	\includegraphics[width=\columnwidth]{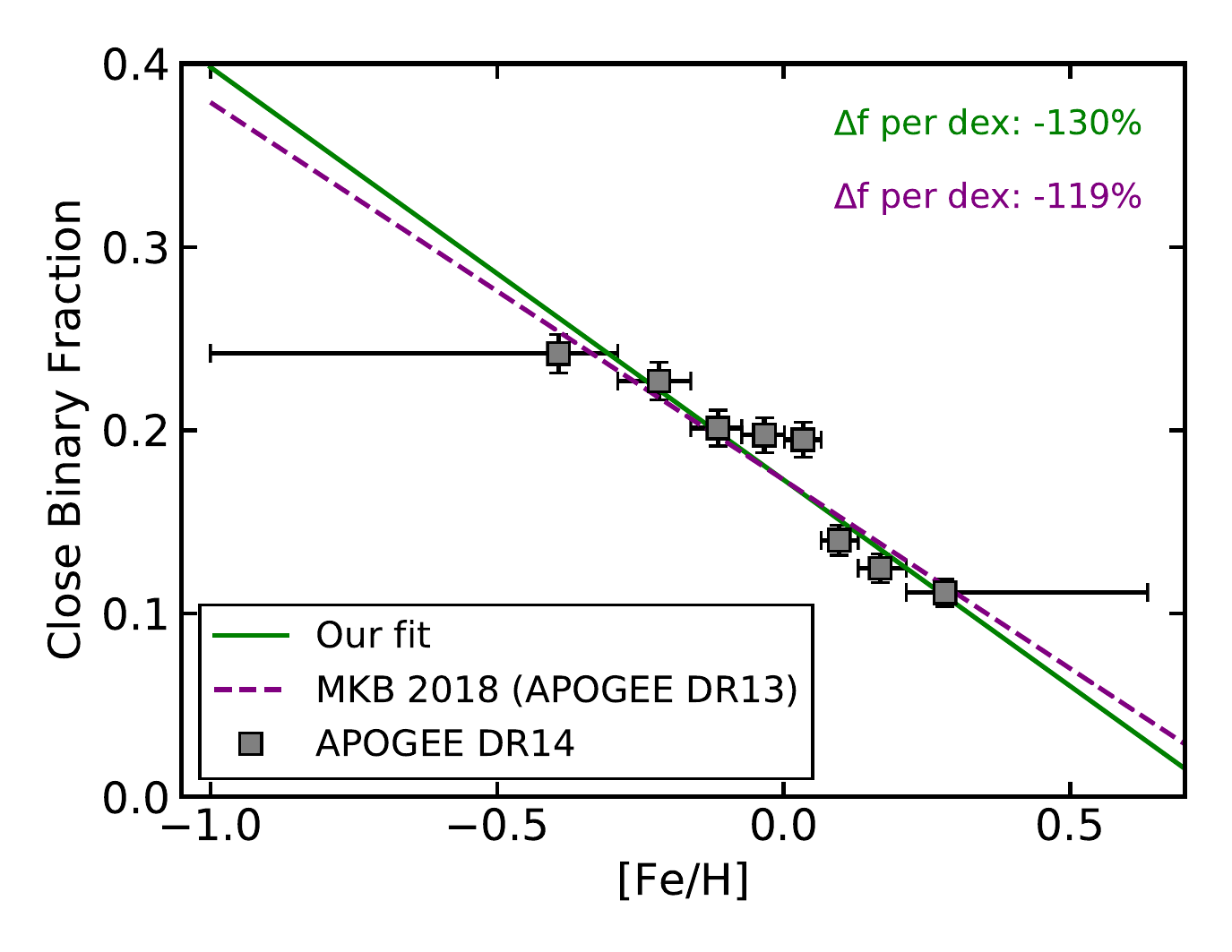}
    \caption{Completeness-corrected close binary fraction for each \feh{} bin. The horizontal error bars show the \feh{} range of each bin, and the vertical error bars show the completeness-adjusted uncertainty, $\sigma_{f}/c$. The results of \citet{Moe2019} are over-plotted alongside a linear fit to our data in order to find the difference in the close binary fraction per dex of \feh{}.}
    \label{fig:fehslope}
\end{figure}

The completeness-corrected close ($\log(P/\text{days})\leq4.0$) binary fraction as a function of \feh{} alone is shown in Fig.~\ref{fig:fehslope}. We divided our sample into eight bins in \feh{}, chosen to contain approximately 5200 stars each. A linear fit to these data shows that the close binary fraction decreases by a factor of $\sim2.4$ from \feh\ $=-0.5$ dex to \feh\ $=0.25$ dex. As we have seen, however, the relationship between chemical composition and stellar multiplicity is complex, and it cannot be characterised by metallicity alone. Here we consider in detail four parameters related to the abundance of $\alpha$-process elements: \mgh{} and \sih{} (already discussed in Section~\ref{subsec:triangle}), plus \alh{} and \oh{}. The measurements of [C/H] and [N/H] for APOGEE DR14 dwarfs are not reliable \citep{Holtzman2018}, so we did not include them in our analysis.

\begin{figure*}
	\includegraphics[width=\textwidth]{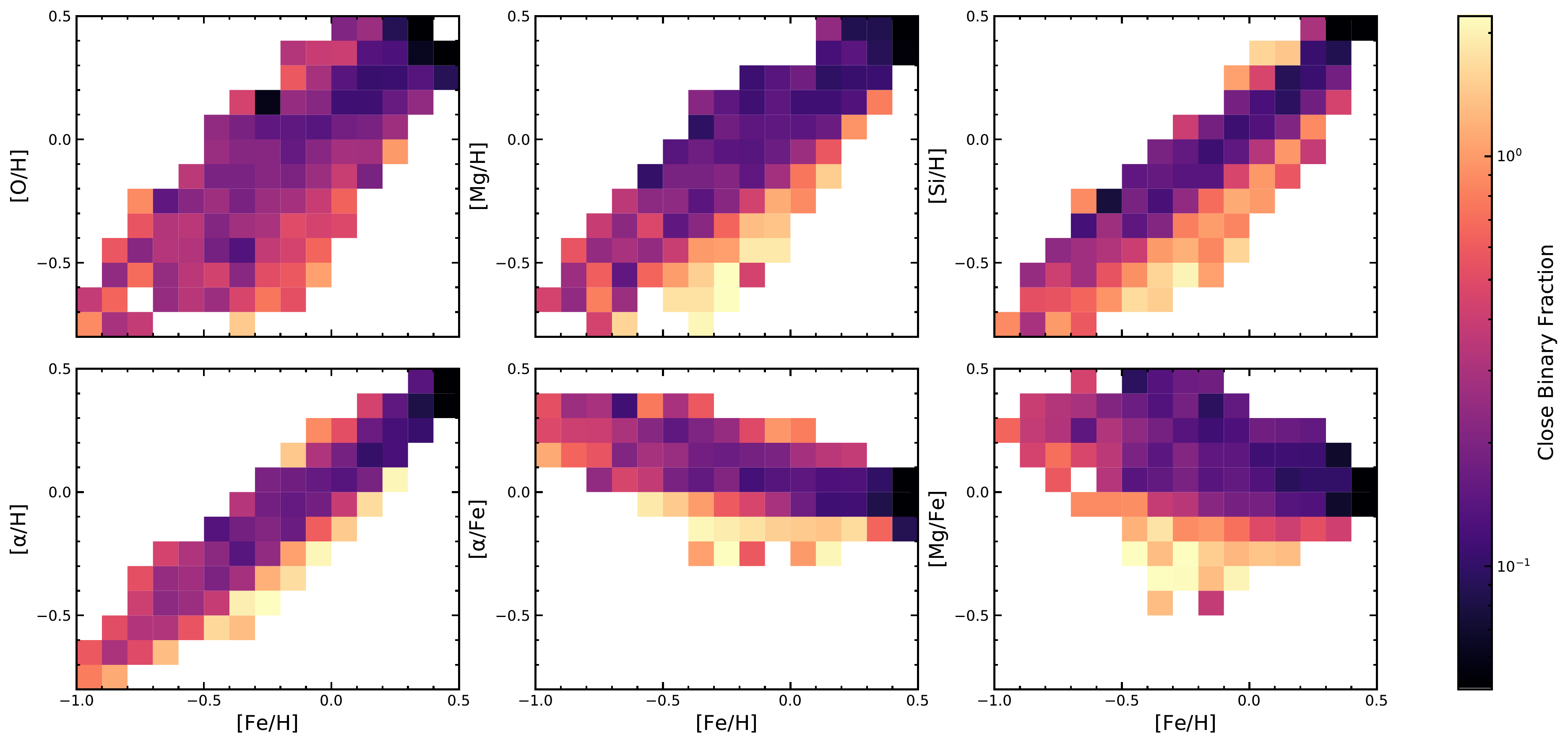}
    \caption{Two-dimensional histogram showing the completeness-corrected close binary fraction as a function of \feh{} and various $\alpha$ abundances.}
    \label{fig:2dfracs}
\end{figure*}

We begin by revisiting the two-dimensional histograms of completeness-corrected close binary fraction. Each panel in Fig.~\ref{fig:2dfracs} shows an $\alpha$-process abundance measurement a function of \feh{}, similar to the 2D histograms of Fig.~\ref{fig:triangle}, but with a lower minimum count of five stars per bin to maximise parameter space coverage. The close binary fraction again exceeds 100 per cent in multiple bins, though this still may due to the degeneracies present in our RV variability fraction method discussed in Sec.~\ref{subsec:triangle}. The anti-correlation between close binary fraction and \feh{} is apparent as the trend along the diagonal, and it is present for all six $\alpha$ abundances. The additional anti-correlation with $\alpha$ abundance is clear when looking along lines of constant \feh{}, manifesting as two distinct sequences: $\alpha$-poor with large close binary fractions, and $\alpha$-rich with smaller close binary fractions. The weakest effect is seen in \oh{}, but the anti-correlation is obvious for \alh{}, \mgh{}, [Mg/Fe], and Si. However, especially around solar metallicity, \alh{}, \oh{}, and \sih{} show increased close binary fractions at low \textit{and} high values when looking along lines of constant \feh{}.

\begin{figure*}
	\includegraphics[width=1.0\textwidth]{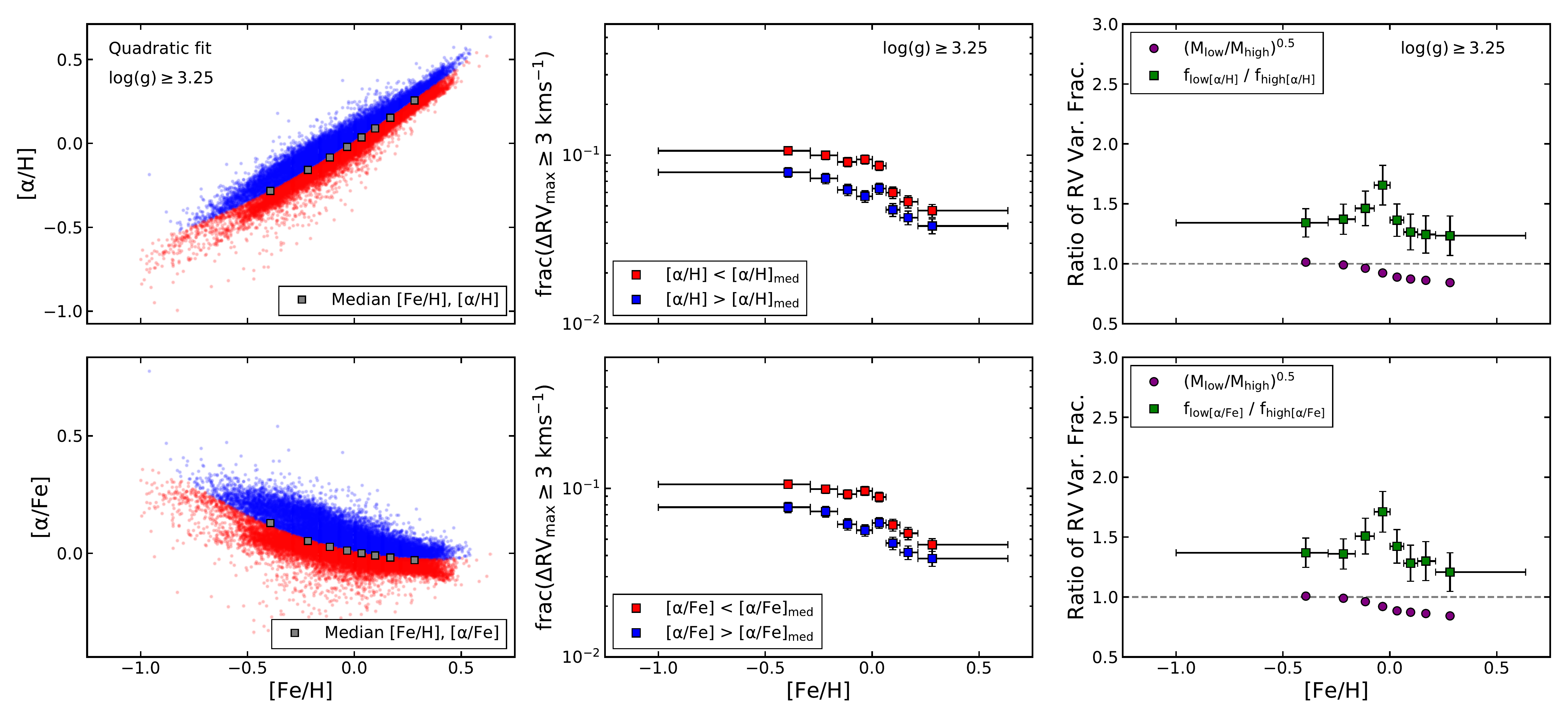}
    \caption{First row: [left] distribution of \alh{} and \feh{}, with the grey points placed at the median \feh{}/\alh{} of each bin and the red points showing the ``low-\alh{}'' subsample and blue the ``high-\alh{}'' subsample; [center] the fraction of systems with \drvm{}$\geq3$ \kms{} for the low- and high-\alh{} subsamples, with the horizontal error bars showing the range of \feh{} in each bin and the vertical error bars showing the uncertainty, equation~(\ref{eq:PN}); and [right] the ratio of the low-$\alpha$ to high-$\alpha$ bins' RV variability fraction alongside the ratio of median masses between the low-$\alpha$ and high-$\alpha$ bins. The second row is the same but for \alfe{}.}
    \label{fig:fracsalphs}
\end{figure*}

\begin{figure*}
	\includegraphics[width=1.0\textwidth]{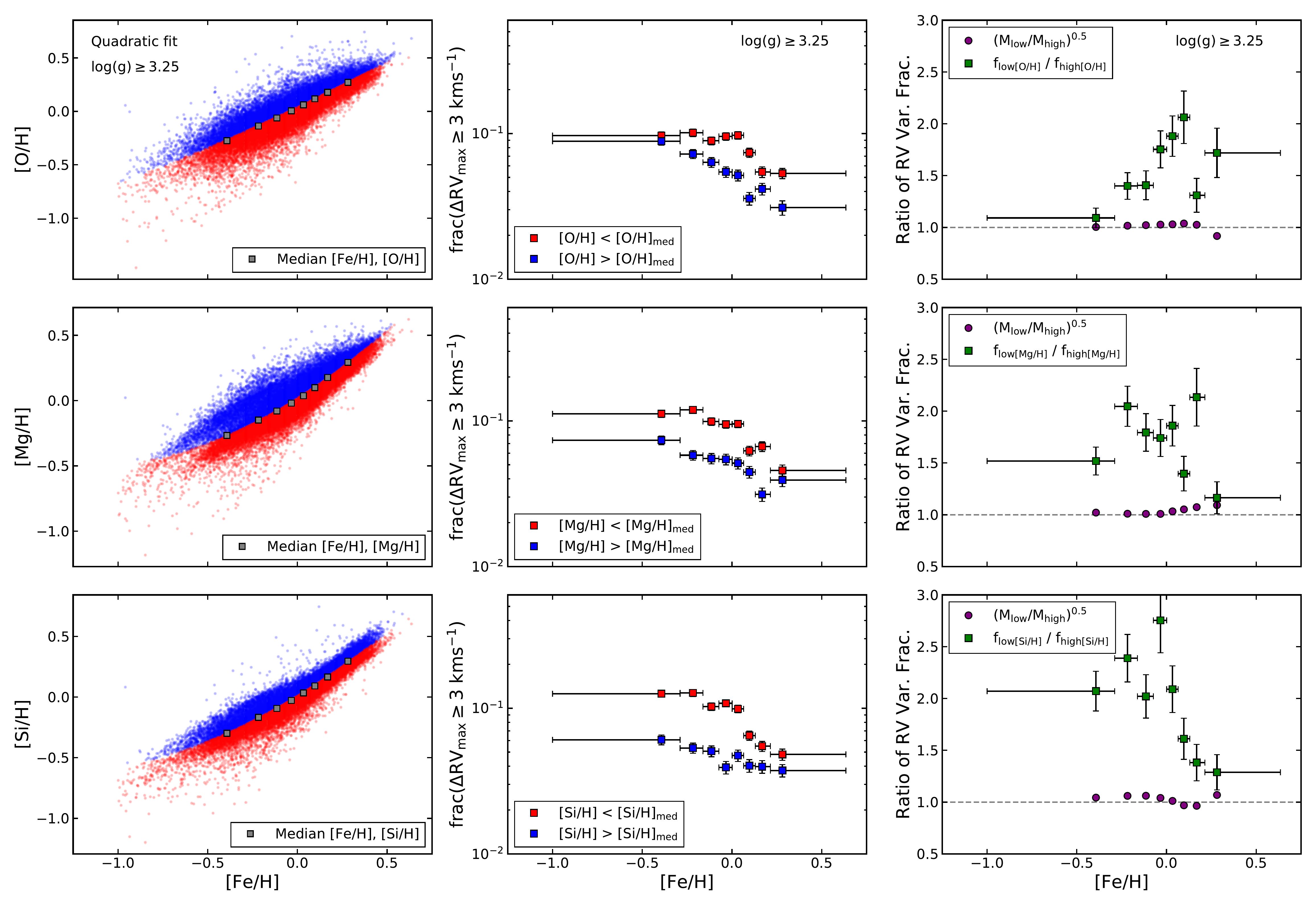}
    \caption{Same as Fig.~\ref{fig:fracsalphs}, but for \oh{}, \mgh{}, and \sih{}.}
    \label{fig:fracsOMgSi}
\end{figure*}

\begin{figure}
	\includegraphics[width=\columnwidth]{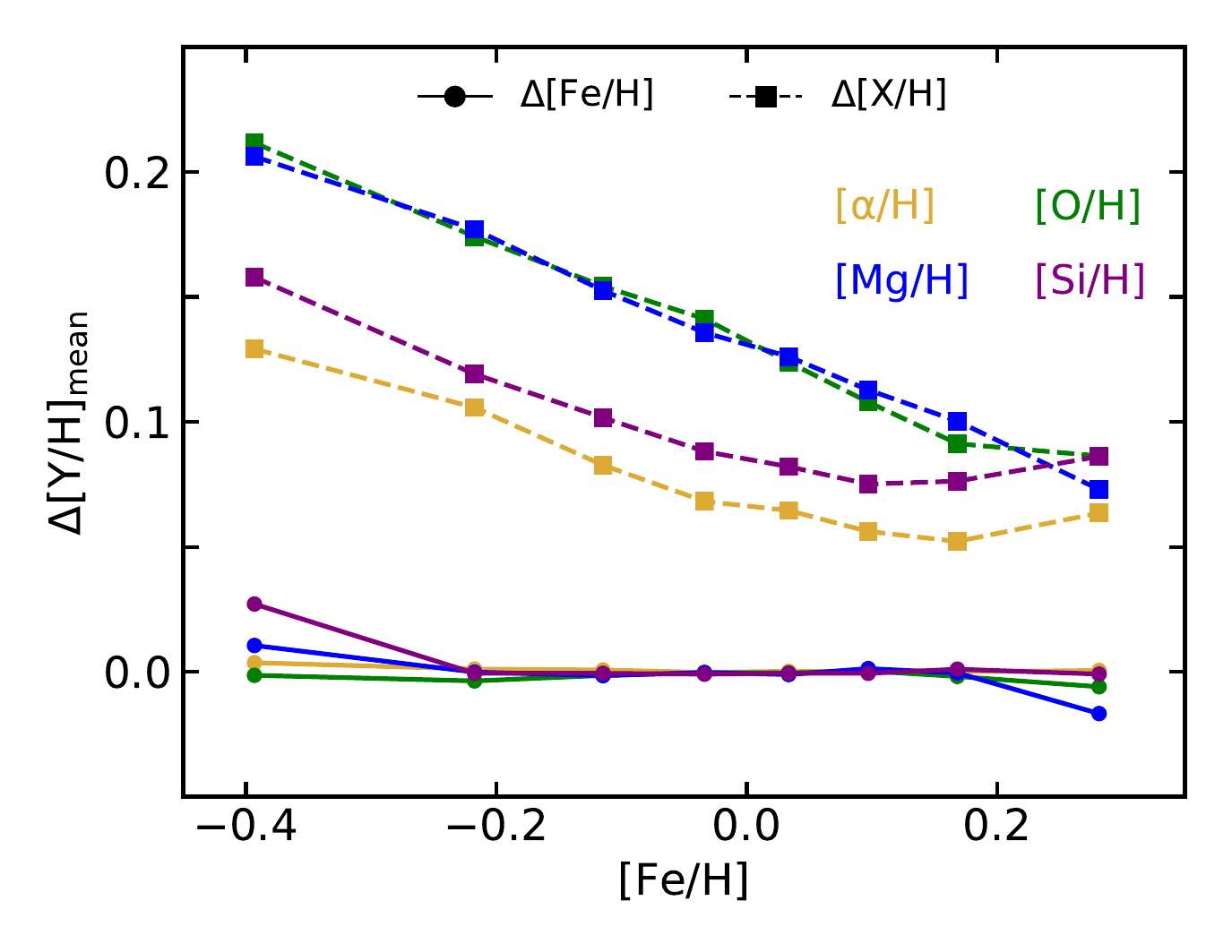}
    \caption{Differences in the mean [Y/H] for each sample, with $\Delta$[Y/H]$_{\text{mean}}$ = mean([Y/H]$_{\text{high}})-$mean([Y/H]$_{\text{low}}$).}
    \label{fig:xhdiffs}
\end{figure}

\begin{figure*}
    \includegraphics[width=\textwidth]{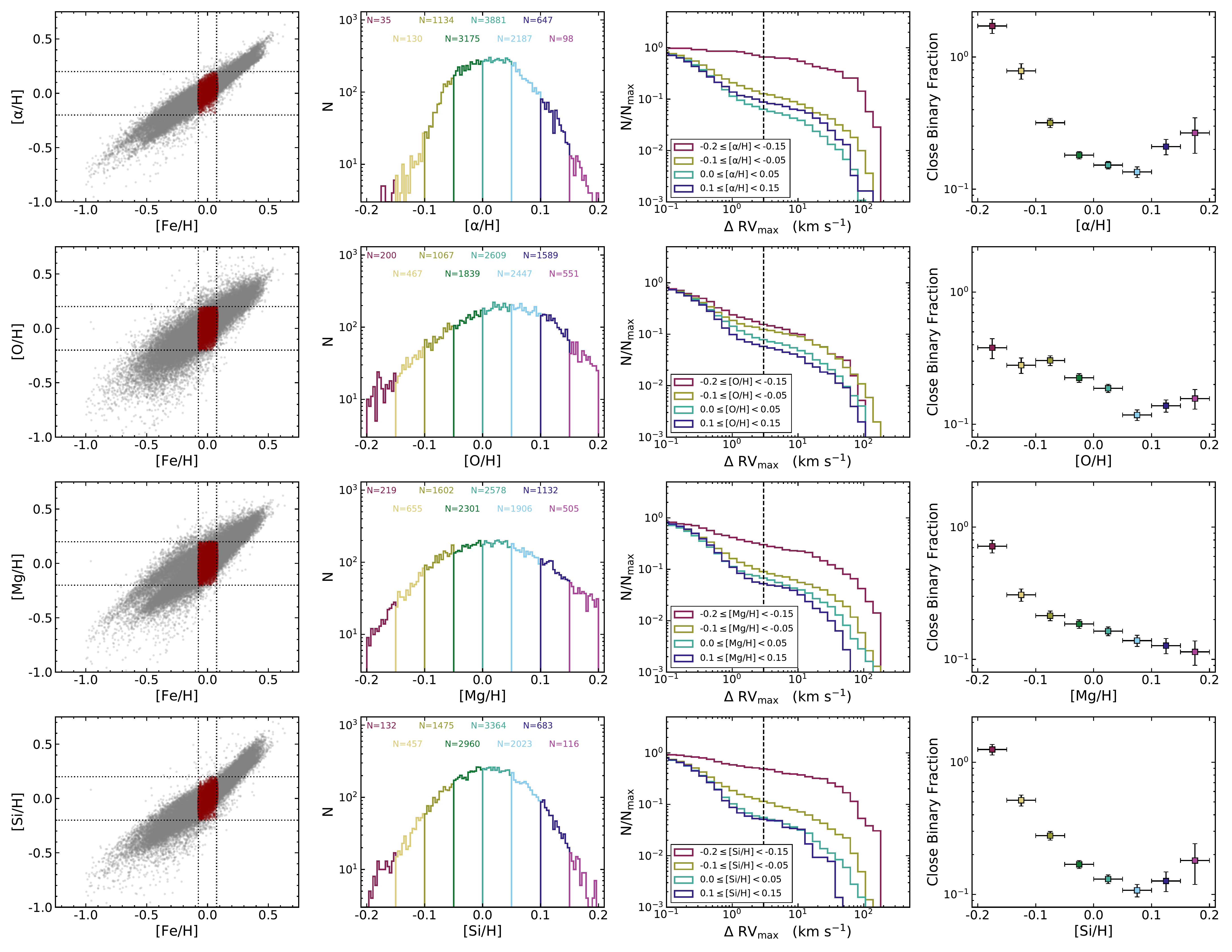}
    \caption{Various distributions for a selection of data in a narrow range around solar \feh{}. First row: [far left] \alh{} versus \feh{}, where the entire sample is shown in grey and the chosen subsample is shown in dark red (boundaries $-0.075\leq$ \feh{} $\leq0.075$ and $-0.2\leq$ \alh{} $\leq0.2$); [center left] histograms for the selected data, split into eight equally spaced bins across \alh{}; [center right] selected cumulative \drvm{} histograms; and [far right] the close binary fraction as a function of \alh{} for the selected data, colour-coded by its \alh{} bin. The horizontal error bars show the \alh{} range of each bin, and the vertical error bars show the completeness-adjusted uncertainty, $\sigma_{f}/c$. The remaining rows are the same but for \oh{}, \mgh{}, and \sih{}.}
    \label{fig:solfehalphas}
\end{figure*}

To study these effects in a regime that is not prone to numerical noise due to small numbers of stars, we use the same bins as those shown in Fig.~\ref{fig:fehslope} ($N\sim5200$ each). The grey squares in the first column of Figs.~\ref{fig:fracsalphs}-\ref{fig:fracsOMgSi} are plotted at the median \feh{} and $\alpha$ abundance for each bin, with each row showing one of the four $\alpha$ abundance measurements from earlier, plus \alfe. Within each \feh{} bin we define ``low-X'' and ``high-X'' subsamples, with X standing for each of the five parameters we study, shown in red and blue. The dividing line between low- and high-X samples is drawn using a quadratic fit to the median with a finer grid of thirty-five bins in \feh{}. The second column of Figs.~\ref{fig:fracsalphs}-\ref{fig:fracsOMgSi} shows the fraction of systems with \drvm{}$\geq3$ \kms{} for each of the low- and high-X subsamples as a function of \feh{}. Horizontal error bars indicate the extent of the \feh{} bins, and vertical error bars represent the binomial process uncertainty, $\sigma_{f}$ (equation~\ref{eq:PN}). The anti-correlation between \feh{} and close binary fraction is again present, but there is a significant gap between the low- and high-X subsamples in all five parameters we study. The green points in the third column of Figs.~\ref{fig:fracsalphs}-\ref{fig:fracsOMgSi} display the ratio of RV variability fractions between the low- and high-X abundance subsamples, with horizontal error bars again indicating the extent of the \feh{} bins and vertical error bars denoting the uncertainty obtained via error propagation. This ratio is greater than one across every \feh{} bin and for every abundance considered here. The ratios generally increase with \feh{} for O, but the opposite appears to be true for Mg and Si.

These results reinforce our finding that $\alpha$ element abundances have a strong impact on the close binary fraction. To further investigate to what extent this effect is separate from the \feh{} effect, we calculated the difference in subsample means for four of the abundance measurements, defined as
\begin{equation}
	\Delta\text{[Y/H]}_{\text{mean}} = \text{mean}(\text{[Y/H]}_{\text{high}}) - \text{mean}(\text{[Y/H]}_{\text{low}})
\end{equation}
where Y can be Fe or one of the $\alpha$-process abundances. We plot these differences in Fig.~\ref{fig:xhdiffs}, which shows that the differences in mean \feh{} are essentially 0 for all of the bins except the first and eighth, while the differences in mean [X/H] for $\alpha$, O, Mg, and Si remain substantial, although they do decrease as \feh{} increases. In other words, while some systematic differences in \feh{} exist between the high-X and low-X samples that we have defined for the $\alpha$ element abundances, they are too small to account for the effect that we see. Of course, our high-X and low-X samples are not exactly comparable in every aspect, but the effect we observe is too large to be due to other (i.e., non-chemistry related) factors. To illustrate this, we also show in the third column of Figs.~\ref{fig:fracsalphs}-\ref{fig:fracsOMgSi} the magnitude of the effect due to systematic differences in the stellar mass measured by \cite{Sanders2018} between the high-X and low-X samples as a function of \feh{}. These systematic differences, while present, are again too small to explain the disparity in RV variability fraction between the low- and high-$\alpha$ subsamples.

Another way to disentangle the \feh{} and $\alpha$ effects is to examine trends with $\alpha$ abundances in a narrow range of \feh{} (Fig.~\ref{fig:solfehalphas}). For each $\alpha$ abundance, we select a subset of the full sample that spans $-0.075\leq$ \feh{} $\leq0.075$ and $-0.2\leq$ [X/H] $\leq0.2$, shown in dark red in the first column of Fig.~\ref{fig:solfehalphas}. We then divide this subsample into eight bins across the relevant $\alpha$ abundance, with histograms for each bin shown in the second column and the number of objects in each bin listed in the coloured text. Cumulative \drvm{} histograms for four of these bins are shown in the third column. The fourth column displays the completeness-corrected close binary fraction as a function of each $\alpha$-process abundance measurement, with the horizontal error bars indicating the edges of the bins and the vertical error bars representing the completeness-adjusted uncertainty, $\sigma_{f}/c$. This analysis reveals that the close binary fraction in this narrow \feh{} range is clearly anti-correlated with Mg. For the other three abundances, there is a general downward trend, but the detailed behaviour is more complex. For \alh{}, O, and Si, it appears to reach a minimum around 0.075 dex, and then steadily increases once again. This turnaround is weakly present in O, but it is clear in \alh{} and Si. Similar to the third panel of Figs.~\ref{fig:fracsalphs}-\ref{fig:fracsOMgSi}, we compared the ratio of the median mass for each bin against the bin with the minimum close binary fraction. Across all abundances, the difference in median mass between bins is insignificant compared to the difference in observed RV variability.

We repeated this analysis for other narrow ranges of \feh{}. Each subsample spanned a width of $\Delta$\feh{} $=0.15$ dex ($\pm0.075$ from the central value) and $\Delta$[X/H] $=0.4$ dex ($\pm0.2$ from the central value). The central values we compared, in pairs of (\feh{}, [X/H]), were ($-0.4$, $-0.3$); ($-0.2$, $-0.2$); and ($0.2$, $0.2$) - the figures for each set are included in Appendix~\ref{appx:extra}. The turnaround that we see in the solar \feh{} sample was present in some, but not all the \feh{} bins. These results confirm the trends seen in Fig.~\ref{fig:2dfracs} with larger sample sizes, and show that the multiplicity statistics for stars with a specific chemistry can be quite extreme - see for instance the prominent tail in the \drvm{} distributions corresponding to the lowest $\alpha$ abundances in Fig.~\ref{fig:solfehalphas}. In these extreme cases, it is possible that our assumed underlying period distribution is incorrect, which would make our derived values of the completeness corrected close binary fractions incorrect. However, our reported high fractions of RV variability are robust, and clearly require a high frequency of close binary companions, regardless of the underlying period distribution.

Finally, we note that when looking along lines of constant $\alpha$ abundance in Fig.~\ref{fig:2dfracs}, the binary fraction is often \textit{positively} correlated with \feh{}. To verify that this is a real effect and not just a result of binning and small number statistics, we examined the cumulative chemistry distributions for both our entire sample and just the objects with \drvm{}$\geq3$ \kms{}. For every $\alpha$ abundance measurement, the cumulative [X/H] distributions across each of the eight \feh{} bins used in Figs.~\ref{fig:fracsalphs}-\ref{fig:fracsOMgSi} reveal that the RV variables are always shifted towards lower $\alpha$ abundances compared to the total population. Again, this confirms the general anti-correlation between close binary fraction and $\alpha$ abundances. We then plotted the cumulative \feh{} distributions for six bins of equal width across $-0.8\leq$ [X/H] $<0.4$. For the first three bins ($-0.8\leq$ [X/H] $<-0.2$), the RV variables are shifted towards \textit{higher} \feh{} abundances than the total population. This is true for all $\alpha$ abundances considered here. However, the cumulative metallicity distributions for the two bins between $-0.2\leq$ [X/H] $<0.2$ generally show a weakening of this trend, and the bin for $0.2\leq$ [X/H] $<0.4$ shows a reversal--i.e., the RV variables are shifted towards \textit{lower} \feh{} than the greater population. These results lend support for a inflection point in the close binary fraction as a function of $\alpha$ abundances around 0.1 dex. 

To summarise our findings, we find a robust anti-correlation between $\alpha$ abundances and close binary fraction, similar in strength but separate from the already established anti-correlation with \feh{}. The general trend of increasing $\alpha$ abundances to decrease the close binary fraction is robust, but the details are complex, and it is likely that the effect is not completely independent from \feh{}, at least in some regimes. Because of this, it might not be possible to provide a simple quantitative description of the full relationship between stellar chemistry and close binary fraction in the APOGEE sample.

\section{Discussion}
\label{sec:discuss}

In Section~\ref{sec:results} we have shown that the relationship between stellar multiplicity and stellar parameters is quite complex. A robust physical interpretation of the observed anti-correlation between $\alpha$-process abundances and close binary fraction thus requires careful consideration of potential systematics and internal correlations. In this section, we address two such effects that were not discussed in Sections~\ref{subsec:triangle} and \ref{subsec:chemCBF}, and we consider the implications that our results have for star formation.

\subsection{Potential Systematics}
\label{subsec:systematics}

\textbf{Visit Histories} Most of our targets have sparsely sampled RV curves. Among the non-SB2s, 36 per cent of objects have only two visits, and 43 per cent have three. For the suspected SB2s, 43 per cent, 35 per cent, and 22 per cent having 2, 3, and 4+ visits, respectively. This is expected; as discussed in Sec.~\ref{sec:samp}, fitting a single stellar template to an SB2 can bias the fit parameters \citep{El-Badry2018a} and also result in poorer fits overall, which are then flagged in the various APOGEE bitmasks. Because we make quality cuts on these bitmasks, we expect that fewer SB2 visits might pass our quality cuts than the overall sample, though we emphasize that our stringent cuts in S/N are still in place. In both cases, objects with duplicate \verb+allStar+ entries (as discussed in the first paragraph of Sec.~\ref{sec:samp}) are biased towards more visits ($\sim$65 per cent with 4+ visits) and longer baselines than non-duplicated APOGEE IDs. For each \feh{} bin used in Figs.~\ref{fig:fracsalphs}-\ref{fig:fracsOMgSi}, we compared the low- and high-$\alpha$ subsamples across histograms of the baselines, JD$_{N}-$JD$_{1}$; the median of the time lags between visits for each star, median(JD$_{i+1}-$JD$_{i}$); and the mean of the time lags between visits for each star, mean(JD$_{i+1}-$JD$_{i}$), where JD is the Julian date of each observation for a star with $N$ total visits. There does not appear to be any significant variation in these parameters with \feh{} or $\alpha$ abundances. The fractions of stars with 2, 3, and 4+ visits for each \feh{} bin and low- and high-$\alpha$ subsample are also consistent with those for the entire sample.

\textbf{White Dwarf Pollution} Some portion of our sample may be post-common envelope systems with white dwarf companions, rather than two MS stars or a subgiant-MS pair. The fraction of these systems will vary with the age of the stellar population, but for short-period ($\log(P/\text{days})\leq4$), it is $\sim15$ per cent at 1 Gyr and $\sim30$ per cent at 10 Gyr \citep[][see their Section 8.3 and Fig.~29]{Moe2017}. Considering our median sample age $\tau\sim8$ Gyr, we expect a fraction of roughly 25 per cent white dwarf companions in our sample. This fraction will also depend upon the metallicity of the stars, but it cannot explain the factor of 1.5-2 difference we see in the close binary fractions of high-$\alpha$ and low-$\alpha$ samples.

\subsection{Implications for Binary Star Formation}
\label{subsec:sfimp}

Close binaries ($a$ $<$ 10 au) likely formed via fragmentation, accretion, and inward migration in the disk, whereas wide binaries ($a$ $>$ 200 au) probably formed via fragmentation of molecular cores \citep{Fisher2004,Kratter2016,Moe2017,Tokovinin2017,Tokovinin2020}. Fragmentation of molecular cores is relatively insensitive to opacity \citep{Bate2014}, explaining why both the initial mass function \citep{Kroupa2013} and wide binary fraction beyond a > 200 au \citep{Moe2019,El-Badry2018} are metallicity invariant across $-$1.0 < \feh{} < 0.5. A natural consequence of such a model is that the close binary fraction increases with primary mass because massive protostellar disks are more prone to fragmentation \citep{Kratter2006}. Analytical models and hydrodynamic simulations also show that the propensity for disk fragmentation decreases with metallicity due to two compounding effects \citep{Machida2009,Tanaka2014,Moe2019}. First, optically thin cores on large spatial scales radiate via molecular transitions, and so metal-poor cores are systematically hotter and must achieve higher masses in order to collapse into disks. The systematically higher core masses toward lower metallicities do however lead to higher accretion rates onto the disks, promoting gravitational instability \citep{Machida2009}. Second, for solar abundances, protostellar disks massive enough to undergo gravitational instability are optically thick \citep{Rafikov2005,Clarke2009,Kratter2010b}. Decreasing the disk's metallicity decreases its optical depth, allowing the mid-plane to radiate and cool more effectively, stimulating disk fragmentation \citep{Tanaka2014,Moe2019}. Note that \citet{Bate2019} has posited a more complex explanation for the increased close binary fraction at low metallicity. While some increase in disk fragmentation is observed in the simulations, \citet{Bate2019} also observes that metal poor cores fragment on very small scales, where the gas is also optically thick. Moreover, due to the very high rate of dynamical interactions observed in these simulations, far more interchanges between core fragmentation and disk fragmentation binaries are observed. The initial conditions in such simulations may not be representative of lower density star clusters in the solar neighbourhood. 

We confirm that the close binary fraction decreases with \feh, consistent with previous observational surveys and theoretical models. Moreover, we demonstrate for the first time that the close binary fraction decreases more rapidly with $\alpha$ than Fe for \alfe{} $<$ 0.05 dex, consistent with expectations from the two compounding effects described above. For example, optically thin cores radiate mainly through molecular CO transitions, and so the infall rates onto the disk are mainly set by $\alpha$ abundances. In the cold ($T<150K$) midplane of disks prone to fragmentation, opacities are dominated by dust and in particular ice covered grains, which can comprise roughly $60$ per cent of the solid particles volume; refractory organics are the second most important contributor in this regime \citep{Semenov2003}. While the optical properties of grains still depend on the distribution and topology of Fe, the changing abundances of O and Si will play a larger role in the bulk opacity. The disk's temperature profile and probability of fragmentation is therefore more dependant on $\alpha$ abundances, explaining why the close binary fraction is anti-correlated with O and Si to a larger degree than with Fe.

For \alfe{} $>$ 0.05 dex, a different picture emerges whereby the close binary fraction within $a$ $<$ 10 au flattens to 10 per cent, independent of chemical abundance. This ``floor'' of a 10 per cent close binary fraction appears to be universal. For example, although the close binary fraction increases from 15 per cent for K-dwarfs to 30 per cent for A-dwarfs, the close binary fraction of both M-dwarfs and brown dwarfs is 10 per cent, relatively constant across $M_1$ = 0.05\,-\,0.6\,M$_{\odot}$ \citep{Joergens2008,Moe2017,Murphy2018,Winters2019,Moe2019b}. One possible explanation is that at least 10 per cent of protostellar disks become massive or cool enough to fragment early in their accretion evolution, regardless of their chemical composition or the final primary mass. Another possibility is that metal-rich and/or low-mass disks are entirely unsusceptible to fragmentation, and the floor of a 10 per cent close binary fraction is actually due to the small fraction of cores that fragment on large scales and subsequently decay to $a$ $<$ 10 au via dynamical friction or exchange interactions \citep{Lee2019,Bate2019}. In the future, measurements of how the close binary fraction of M-dwarfs changes with Fe and $\alpha$ will help differentiate between these two scenarios. 

Another consequence of these metallicity trends is that the overall companion distribution becomes skewed towards shorter separations with decreasing metallicity (see Fig.~19 in \citealt{Moe2019}). However, based on the DR13 sample of SDSS-APOGEE RV variables and {\it Kepler} eclipsing binaries, \citet{Moe2019} found that the separation distribution of solar-type binaries across $a$ = 0.02\,-\,10 au does not vary with metallicity at a statistically significant level. With our larger DR14 sample of SDSS-APOGEE RV variables, we find that $\alpha$-poor binaries are skewed toward larger \drvm{} and thus shorter separations. With decreasing metallicity, models suggest that disks are not only more likely to fragment, but that disk fragmentation occurs at smaller separations \citep{Machida2009,Moe2019}. For example, Fig.~10 of \citet{Machida2009} shows that fragmentation occurs near 200 au at solar-metallicity but near 1 au for Population III stars. Similarly, according to Fig.~20 of \citet{Moe2019}, disks with solar-metallicity are stable within $a$ $<$ 30 au, but metal-poor disks with Z = 10$^{-3}$ Z$_{\odot}$ are capable of fragmentation near $a$ = 8 au. Our measurements are qualitatively consistent with these models, demonstrating that $\alpha$-poor stars not only have a higher close binary fraction, but that those close binaries are skewed toward shorter separations. Note that at very low metallicities, one might expect the relative importance of Fe vs $\alpha$ elements to shift; if fragmentation is pushed to closer separations and correspondingly higher temperatures, the relative importance of ices and organics decreases, and the overall iron abundance might become more important \citep{Semenov2003}.

\section{Conclusions}
\label{sec:concl}

We have presented an analysis of the complex relationship between stellar multiplicity and stellar parameters, with an emphasis on the trends for various $\alpha$-process abundances. We defined a sample of 41,363 dwarf and subgiant stars from APOGEE DR14 with well-measured stellar parameters and at least 2 RV measurements. Because most objects in our sample have sparsely-sampled RV curves, we applied a threshold on the maximum RV shift, \drvm{}$\geq3$ \kms{}, to calculate a fraction of RV variables. This fraction is a tracer for the close binary fraction, modulo a completeness correction that can be estimated for the APOGEE observing epochs using a Monte Carlo method with an assumed period distribution. We analysed these completeness-corrected close binary fractions alongside a variety of stellar parameters: \Teff, $M$, \feh, \mgh, \sih, $\tau$, $J_{z}$, and $v_{R}$. We report a strong anti-correlation between the close binary fraction and Mg and Si abundances, similar in strength but separate from the known anti-correlation with \feh{}. Other stellar parameters like \Teff\ and $M$ also have an impact on the close binary fraction, but chemical composition is clearly the main driver of multiplicity trends in our APOGEE sample.

We further investigated the relationship between \feh{}, $\alpha$-process abundances, and stellar multiplicity, measuring a slightly steeper anti-correlation between \feh{} and the close binary fraction across the narrower interval $-0.4\leq$ \feh{} $\leq0.3$ than the average slope across $-1.0\leq$ \feh{} $\leq0.5$ reported by \citet{Moe2019}, similar to the trend found by \cite{El-Badry2018}. The observed anti-correlations between the close binary fraction and $\alpha$-process abundances ($\alpha$, O, Mg, Si) are \feh{}- and abundance-dependant in strength and consistency. Mg and Si in particular showed exceptionally large close binary fractions and remarkable \drvm\ distributions, where the cores almost disappeared entirely. We also find evidence for a correlation \textit{and} anti-correlation between the close binary fraction and \alh{} and \sih{} with a narrow range of our parameter space. The anti-correlation between stellar composition and close multiplicity fraction has a basis in stellar formation theory. However, low-$\alpha$ binaries are also expected to be skewed towards shorter separations, which would also result in an excess of RV variables independent of an increase in the close binary fraction. Future studies of the period distribution as a function of metallicity and chemistry will help clarify the magnitude of these two effects within our measurements.

\section*{Acknowledgements}
We are grateful to Jennifer van Saders, Jo Bovy, Jeffrey Newman, and Christian Hayes for discussions. This project was presented to the 2019 APOGEE Stellar Companions Sprint, hosted by the University of Virginia Department of Astronomy, and the attendees provided useful feedback and discussions.

C.N.M. acknowledges support from Scialog Scholar grant 24215 from the Research Corporation. C.N.M. and C.B. acknowledge support from the National Science Foundation grant AST-1909022. S.E.K and M.G.W. acknowledge support from the National Science Foundation grant AST-1909584.
M.M. and K.M.K acknowledge support from NASA grant 80NSSC18K0726 and The Research Corporation for Science Advancement grant ID\# 26077. B.A., H.M.L, and S.R.M. acknowledge support from National Science Foundation grant AST-1616636. N.D. would like to acknowledge that this material is based upon work supported by the National Science Foundation under Grant No. 1616684.

Funding for the Sloan Digital Sky Survey IV has been provided by the Alfred P. Sloan Foundation, the U.S. Department of Energy Office of Science, and the Participating Institutions. SDSS acknowledges support and resources from the Center for High-Performance Computing at the University of Utah. The SDSS web site is www.sdss.org.

SDSS is managed by the Astrophysical Research Consortium for the Participating Institutions of the SDSS Collaboration including the Brazilian Participation Group, the Carnegie Institution for Science, Carnegie Mellon University, the Chilean Participation Group, the French Participation Group, Harvard-Smithsonian Center for Astrophysics, Instituto de Astrof\'{i}sica de Canarias, The Johns Hopkins University, Kavli Institute for the Physics and Mathematics of the Universe (IPMU) / University of Tokyo, the Korean Participation Group, Lawrence Berkeley National Laboratory, Leibniz Institut f\"{u}r Astrophysik Potsdam (AIP), Max-Planck-Institut f\"{u}r Astronomie (MPIA Heidelberg), Max-Planck-Institut f\"{u}r Astrophysik (MPA Garching), Max-Planck-Institut f\"{u}r Extraterrestrische Physik (MPE), National Astronomical Observatories of China, New Mexico State University, New York University, University of Notre Dame, Observat\'{o}rio Nacional / MCTI, The Ohio State University, Pennsylvania State University, Shanghai Astronomical Observatory, United Kingdom Participation Group, Universidad Nacional Aut\'{o}noma de M\'{e}xico, University of Arizona, University of Colorado Boulder, University of Oxford, University of Portsmouth, University of Utah, University of Virginia, University of Washington, University of Wisconsin, Vanderbilt University, and Yale University.

\emph{Software acknowledgements:} Astropy \citep[]{Robitaille2013}, Jo Bovy's \verb+apogee+ tools \citep[]{Bovy2016}, Ipython \citep[]{Perez2007}, Matplotlib \citep[]{Hunter2007}

\section*{Data Availability}
The stellar parameters, abundances and RVs from APOGEE DR14 were derived from the \verb+allStar+ file at \url{https://www.sdss.org/dr14/irspec/spectro\_data/}. The \cite{Sanders2018} catalogue can be found at \url{https://www.ast.cam.ac.uk/jls/data/gaia_spectro.hdf5}, and details for downloading our likely SB2s are provided in Appendix~\ref{appx:sb2s}. Data from the MC sampler are available from C.N.M. upon reasonable request.




\bibliographystyle{mnras}
\bibliography{alpha_elems_paper_refs_v4}



%
%

\appendix

\section{Choices for the Monte Carlo Simulation}
\label{appx:MC}

We altered several of the choices listed in Sec.~\ref{subsec:MCcorrec} to gauge their effects on the completeness estimate. The first is our choice of \logg{}; it only affects the calculations for the critical period $P_{\text{crit}}$ and circularisation period $P_{\text{circ}}$ \citep{Badenes2018}. The critical period is calculated using
\begin{equation}
    P_{\text{crit}} = \frac{2 \pi}{\sqrt{R^{3}(q)(1+q)}} \Big({\frac{GM}{g^{3}}}\Big)^{1/4}
\end{equation}
where $M$ is the mass of the primary in grams, $g$ is the primary's surface gravity in cm s$^{-2}$, $q$ is the system's mass ratio, and $R(q)$ is the ratio between the radius of the Roche Lobe and the orbital separation \citep{Eggleton1983}. The circularisation period is calculated for \loggunits$=4.25$, $M=1.0$\msun, and \feh=0.0 dex. For $1M_{\odot}$ and \loggunits$=4.25$, we calculate $\log(P_{\text{circ}}/\text{days})=0.888$ and $\log(P_{\text{crit}}/\text{days})=-0.313$. Running the MC with 2x and 0.5x these values resulted in no significant change to the completeness estimate, which is expected; given the \cite{Raghavan2010} period distribution, very few stars will be at short enough periods to be affected by shifts of this magnitude in either of these parameters.

We implemented a 25 per cent increase in twins ($0.95\geq q \geq 1.0$) motivated by the results of \cite{Moe2017}. Reducing this fraction, even to 0 per cent, did not significantly change our completeness estimate.

\begin{figure}
    \includegraphics[width=\columnwidth]{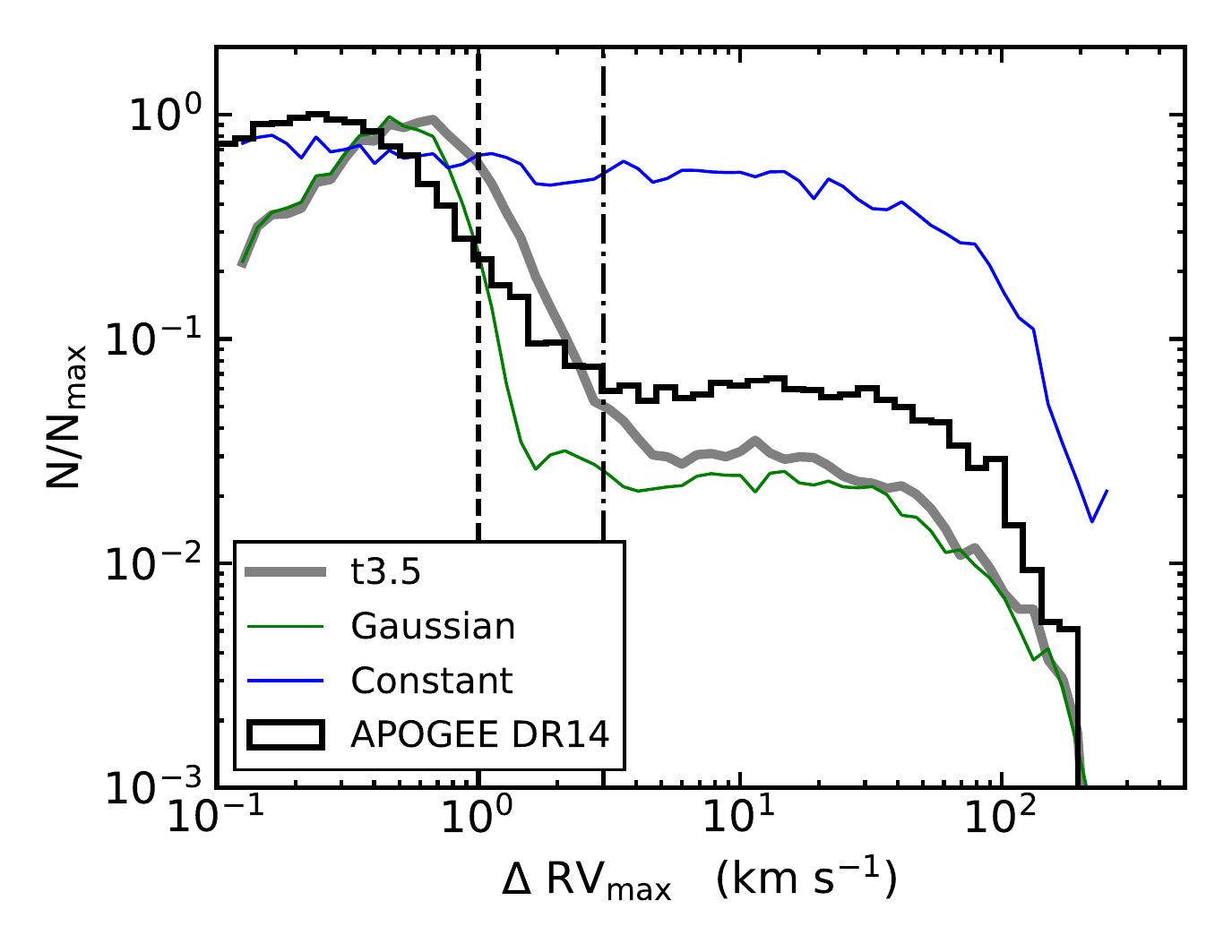}
    \caption{Distributions of \drvm{} from our APOGEE DR14 sample (black) and simulated by our MC with three RV error distributions. All three RV error distributions have $\sigma=0.25$ \kms, and the Gaussian and Student's t with degrees of freedom 3.5 (t3.5) both have $\mu=0$.}
    \label{fig:rverrs}
\end{figure}

Fully characterising the observed RV uncertainty distribution is a difficult task, especially given the diverse nature of our sample. Consequently, our choice for a simulated RV error distribution is not motivated by physical intuition; rather, we have tried a variety of possible distributions and distribution parameters in order to roughly reproduce the shape of the observed \drvm{} distribution. We have found the Student's t distribution to reproduce the relative shape of the core and tail for a variety of \logg{} values, and so we chose to use it in the completeness estimate described in Section~\ref{subsec:MCcorrec}. This choice does not significantly affect our completeness estimates, however. Table~\ref{tab:appa} gives completeness fractions given two \drvm{} thresholds for three RV error distributions:
\begin{enumerate}
    \item a Student's t (scipy.stats.t) with degrees of freedom 3.5, $\mu=0.0$, $\sigma=0.25$ \kms\ (the same used in Section~\ref{sec:results})
    \item a Gaussian (scipy.stats.norm) with $\mu=0.0$, $\sigma=0.25$ \kms\
    \item a constant RV error of $\sigma_{\text{RV}}=0.25$ \kms\ applied to every simulated RV
\end{enumerate}
The simulated \drvm{} distribution for each choice are shown alongside the observed distribution in Fig.~\ref{fig:rverrs}. The completeness fractions are nearly identical between the Student's t and the Gaussian, and while the Student's t has a larger percentage of false positives, it is still a modest increase. The constant RV uncertainty has very similar detection efficiencies to the other two at the low-$P$ end, where the RV variability is the largest. As expected, it diverges for larger periods, where it fails to distinguish between a core and tail in the \drvm\ distribution, apparent in Fig.~\ref{fig:rverrs}. Between the Student's t and Gaussian, the \drvm{} distributions are qualitatively similar. The Gaussian is slightly narrower with a sharper transition between the core and tail, whereas the Student's t has a slightly better match to the overall shape of our observed \drvm{} distribution. For this reason, we chose the Student's t, though we note that this reason is purely qualitative and our choice does not significantly affect the completeness fractions. Future work to better understand the RV uncertainties may favour one distribution over another, but that is beyond the scope of this work.

From theoretical predictions (see Section \ref{subsec:sfimp}) and recent observations of eclipsing binaries \citep{Jayasinghe2020}, the period distribution for solar-type binaries may depend on chemistry. We briefly explored this scenario by implementing a three-component period distribution. Motivated by Fig.~19 of \cite{Moe2019}, we simulated a sample of $N=50,000$ stars using three chemistry-dependant log-normal period distributions:
\begin{enumerate}
    \item $-1\leq$ \feh{} $\leq-0.2$: $\{\mu_{\log(P/\text{days})}=4.0$, $\sigma_{\log(P/\text{days})}=1.5\}$ 
     \item $-0.2\leq$ \feh{} $\leq0.2$: $\{\mu=5.0$, $\sigma=2.0\}$
      \item \feh{} $\geq0.2$: $\{\mu=6.0$, $\sigma=2.5\}$
\end{enumerate}
where \feh\ values are given in dex. We simulated $N_{1}=6,250$, $N_{2}=37,500$, $N_{3}=6,250$, which is proportional to the number of stars in our sample within those \feh\ ranges. The calculated completeness fractions for each MC subsample do not vary significantly from those shown in Table~\ref{tab:compfrac}, and when we combine the subsamples into a single $N=50,000$ sample, the calculated completeness fractions also do not vary significantly. Changing the subsample sizes to $(15,000; 20,000; 15,000)$ also did not result in significantly different completeness fractions.

\begin{table*}
	\centering
	\caption{Completeness and false positive fractions for selected $\log(P/\text{days})$ and \drvm{} thresholds, given three RV error distributions all with $\sigma=0.25$ \kms.}
	\label{tab:appa}
	\begin{tabular}{l|cc|cc|cc}
		\hline
		& \multicolumn{3}{|c|}{\drvm{}$\geq1$ \kms{}} &\multicolumn{3}{|c|}{\drvm{}$\geq3$ \kms{}} \\
		\hline
		$\log(P/\text{days})$ threshold & t3.5 & Gaussian & Constant & t3.5 & Gaussian & Constant\\
		\hline
		$\log{P}\leq0.0$ & 1.00 & 1.00 & 1.00  & 0.96 & 0.97 & 1.00 \\
		$\log{P}\leq2.0$ & 0.93 & 0.94 & 0.93 & 0.84 & 0.84 & 0.85 \\
		$\log{P}\leq4.0$ & 0.55 & 0.50 & 0.64 & 0.34 & 0.34 & 0.47 \\
		\hline
		False Positive Fraction & 2.68\% & 1.20\% & 0.0\% & 0.10\% & 0.0\% & 0.0\% \\

		\hline
	\end{tabular}
\end{table*}

\section{Additional Figures}
\label{appx:extra}

\begin{figure*}
    \includegraphics[width=\textwidth]{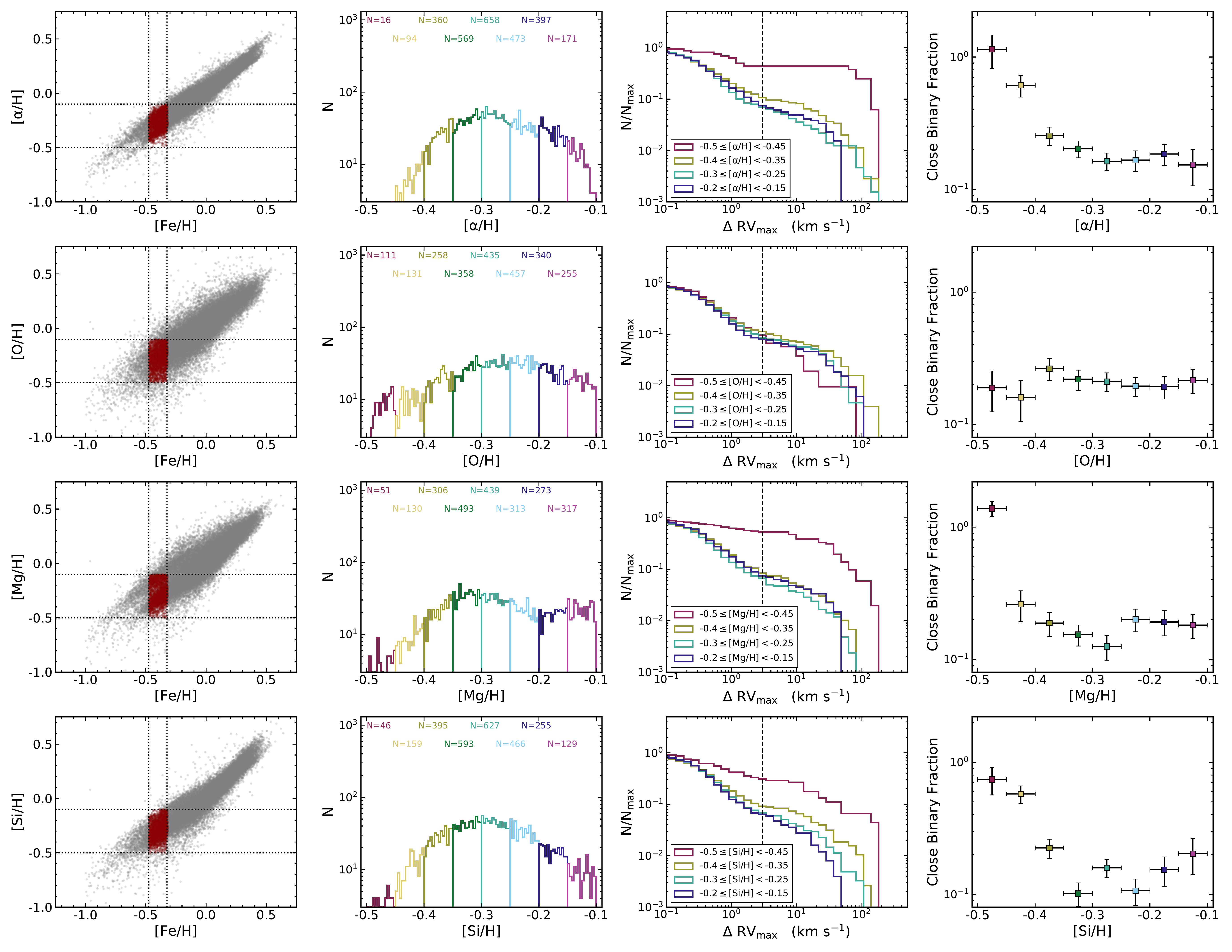}
    \caption{Identical to Fig.~\ref{fig:solfehalphas} but with boundaries $-0.475\leq$ \feh{} $\leq-0.325$ and $-0.5\leq$ X $\leq-0.1$.}
    \label{fig:m04fehalphas}
\end{figure*}

\begin{figure*}
    \includegraphics[width=\textwidth]{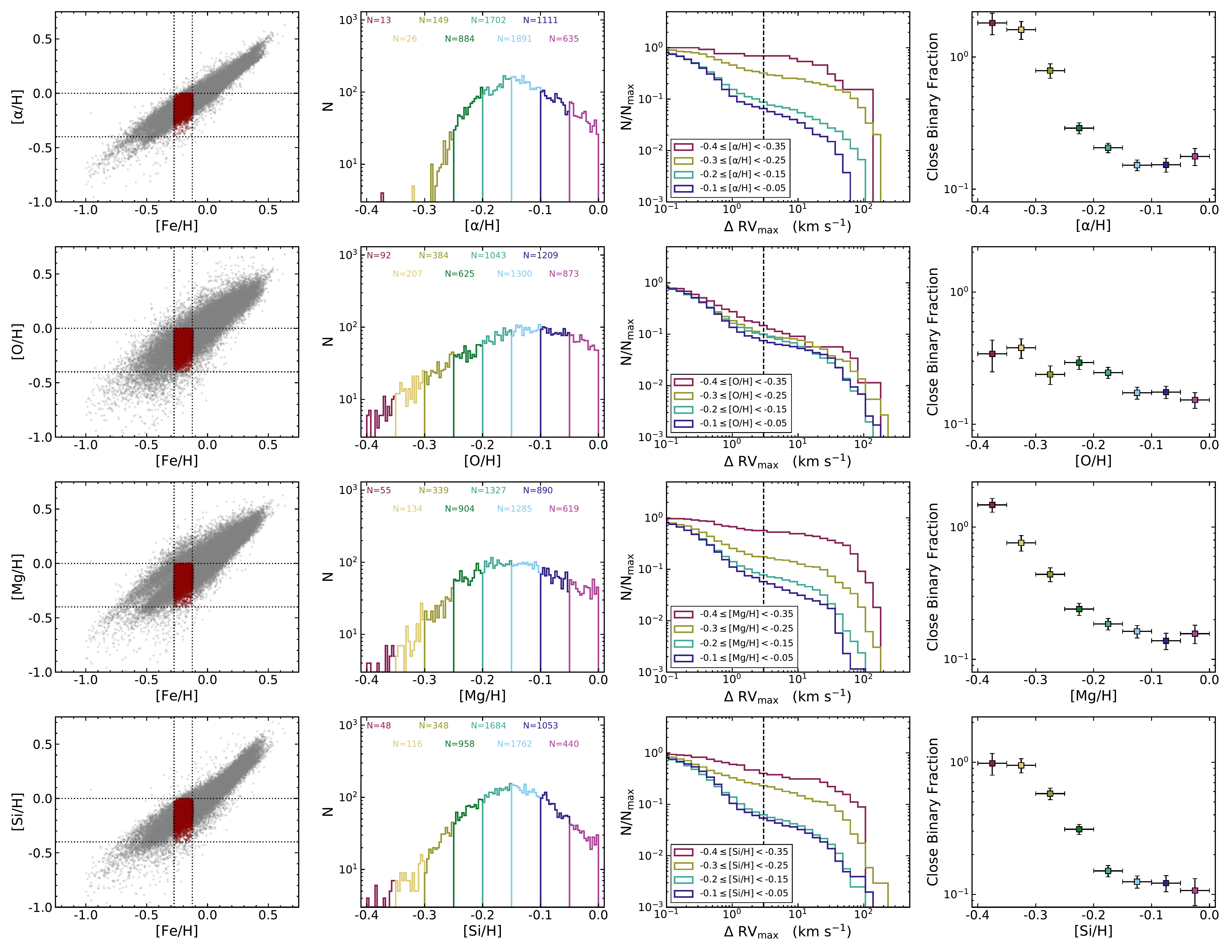}
    \caption{Identical to Fig.~\ref{fig:solfehalphas} but with boundaries $-0.275\leq$ \feh{} $\leq-0.125$ and $-0.4\leq$ X $\leq0.0$.}
    \label{fig:m02fehalphas}
\end{figure*}

\begin{figure*}
    \includegraphics[width=\textwidth]{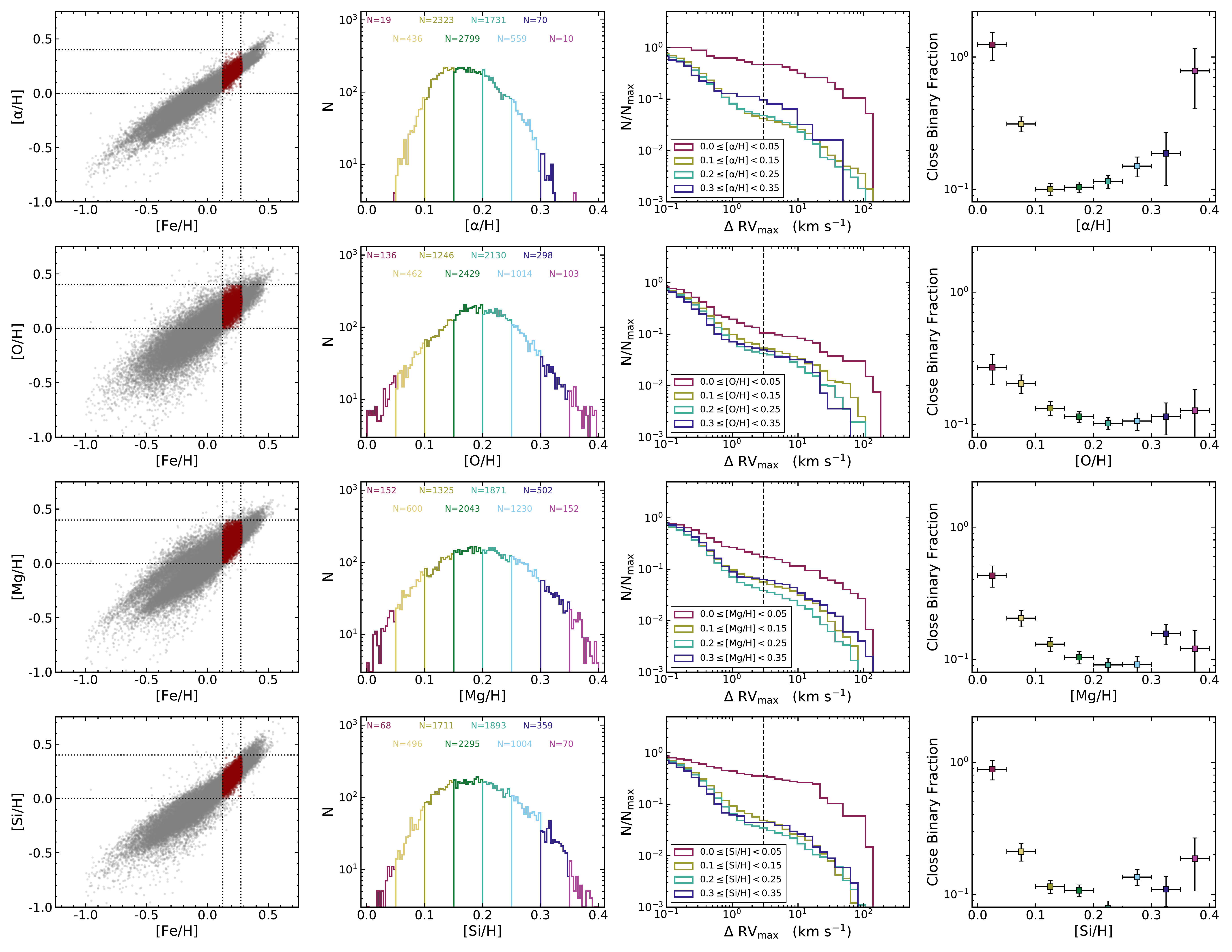}
    \caption{Identical to Fig.~\ref{fig:solfehalphas} but with boundaries $0.125\leq$ \feh{} $\leq0.275$ and $0.0\leq$ X $\leq0.4$.}
    \label{fig:p02fehalphas}
\end{figure*}

\section{Format of Data Products}
\label{appx:sb2s}
Here, we make available the likely SB2s identified by two methods as discussed in Sec.~\ref{sec:samp}. Table~\ref{tab:appxsb2sinfo} describes the column structure, and Tables~\ref{tab:appsb2sap}-\ref{tab:appsb2smk} provide the results from each method. Each entry in Tables~\ref{tab:appsb2sap}-\ref{tab:appsb2smk} is for an individual visit spectrum; the table from re-analysing APOGEE CCFs has 13,970 total entries, and the table from re-calculating the CCFs has 12,044. There are 2832 and 2238 unique APOGEE IDs in each table, respectively.

\begin{table*}
	\centering
	\caption{Format of provided SB2 catalogues. Listed below are the name of each column with a description and any applicable units. For each entry that is given by an array, the array has 8 elements.}
	\label{tab:appxsb2sinfo}
	\begin{tabular}{llc}
		\hline
		Name & Description & Units \\
		\hline
		OBJID & identifier in the APOGEE catalogue & --- \\
		PLATE & APOGEE visit plate ID & --- \\
		FIBER & APOGEE visit fiber ID & --- \\
		MJD & Modified Julian Date of APOGEE visit & --- \\
		N & number of deconvolved components & --- \\
		FLAG & array of integer quality flags, from 1.0 to 4.0 & --- \\
		POS & array of RVs for each component, ordered by amplitude & \kms \\
		AMP & array of amplitudes & --- \\
		FWH & array of full widths at half maximum & \kms \\
		EPOS & array of RV uncertainties & \kms \\
		EAMP & array of amplitude uncertainties & --- \\
		EFWH & array of full width at half maximum uncertainties & \kms \\
		\hline
	\end{tabular}
\end{table*}

\begin{landscape}
\begin{table}
    \caption{Table of likely SB2s identified by re-analysing the APOGEE CCFs. This table is available in its entirety (with parameters for 13,970 visits with 2832 unique APOGEE IDs) in machine-readable form.}
    \label{tab:appsb2sap}
    \begin{tabular}{lccccccccccc}
        \hline
        OBJID & PLATE & FIBER & MJD & N & FLAG & POS & AMP & FWH & EPOS & EAMP & EFWH \\
        \hline
        2M00023036+8524194 & 7950 & 224 & 57295 & 1 & 4.0 ... 0.0 & $-$4.5720005 ... nan & 0.5707569 ... nan & 49.42374 ... nan & 0.57283086 ... nan & 0.013487913 ... nan & 1.3489138 ... nan \\
        2M00023036+8524194 & 9084 & 224 & 57556 & 2 & 4.0 ... 0.0 & $-$71.81711 ... nan & 0.37271407 ... nan & 34.15913 ... nan & 0.546208 ... nan & 0.012155449 ... nan & 1.2862214 ... nan \\
        2M00023179+1521164 & 6560 & 77 & 56584 & 2 & 4.0 ... 0.0 & 20.618723 ... nan & 0.49165606 ... nan & 54.235355 ... nan & 1.8823811 ... nan & 0.014511261 ... nan & 3.0066836 ... nan \\
        2M00023179+1521164 & 6560 & 89 & 56588 & 1 & 4.0 ... 0.0 & 3.9876366 ... nan & 0.54505944 ... nan & 62.63368 ... nan & 0.6144675 ... nan & 0.010904319 ... nan & 1.4469602 ... nan \\
        ... & ... & ... & ... & ... & ... & ... & ... & ... & ... & ... & ... \\
        \hline
    \end{tabular}
\end{table}

\begin{table}
    \caption{Table of likely SB2s identified by calculating our own CCFs. This table is available in its entirety (with parameters for 12,044 visits with 2238 unique APOGEE IDs) in machine-readable form.}
    \label{tab:appsb2smk}
    \begin{tabular}{lccccccccccc}
        \hline
        OBJID & PLATE & FIBER & MJD & N & FLAG & POS & AMP & FWH & EPOS & EAMP & EFWH \\
        \hline
        2M00023036+8524194 & 5095 & 233 & 55821 & 2 & 4.0 ... 0.0 & 50.38858 ... nan & 0.3384734 ... nan & 33.308804 ... nan & 0.51698554 ... nan & 0.010713303 ... nan & 1.217408 ... nan \\
        2M00023036+8524194 & 5095 & 230 & 55824 & 2 & 4.0 ... 0.0 & $-$74.34177 ... nan & 0.32860735 ... nan & 31.727573 ... nan & 0.54551804 ... nan & 0.011521589 ... nan & 1.2845968 ... nan \\
        2M00023036+8524194 & 5095 & 230 & 55840 & 2 & 4.0 ... 0.0 & 66.387726 ... nan & 0.29461184 ... nan & 20.609297 ... nan & 0.94517577 ... nan & 0.027554285 ... nan & 2.2257187 ... nan \\
        2M00023036+8524194 & 5095 & 239 & 55844 & 2 & 4.0 ... 0.0 & 42.22613 ... nan & 0.31082585 ... nan & 24.475822 ... nan & 0.5241617 ... nan & 0.013574529 ... nan & 1.2343063 ... nan \\
        ... & ... & ... & ... & ... & ... & ... & ... & ... & ... & ... & ... \\
        \hline
    \end{tabular}
\end{table}
\end{landscape}


\bsp	
\label{lastpage}
\end{document}